\documentclass[a4paper,11pt]{article}
\pdfoutput=1
\usepackage{jheppub}
\usepackage{longtable,tabu}
\usepackage{array}
\newcolumntype{P}[1]{>{\centering\arraybackslash}p{#1}}
\usepackage{graphicx}
\usepackage{epstopdf}
\usepackage{mathrsfs}
\usepackage{amssymb}
\usepackage{verbatim}
\usepackage{color}
\usepackage[dvipsnames]{xcolor}
\usepackage{multirow}
\usepackage{amsmath}
\usepackage{subfig}
\usepackage[T1]{fontenc}
\usepackage{slashed}
\usepackage{float}
\usepackage[normalem]{ulem}
\usepackage{sidecap}
\usepackage{mathtools}
\usepackage[noabbrev]{cleveref}
\usepackage{diagbox}
\usepackage{lipsum}
\usepackage{caption}
\usepackage[compat=1.0.0]{tikz-feynman}
\usepackage{diagbox}
\usepackage[parfill]{parskip}
\numberwithin{equation}{section}
\numberwithin{figure}{section}
\numberwithin{table}{section}

\begin{document}

\begin{center}
\title{\boldmath
Thermalization in the presence of a time-dependent dissipation and its impact on dark matter production}
\end{center}

\author{Debtosh Chowdhury,}
\author{Arpan Hait}

\affiliation{Department of Physics, Indian Institute of Technology Kanpur, Kanpur 208016, India}
\emailAdd{debtoshc@iitk.ac.in}
\emailAdd{arpan20@iitk.ac.in}


\abstract{In standard cosmological scenarios, a heavy meta-stable field dominates the energy density of the universe after inflation. The dissipation of this field continuously sources high-energy particles. In general, the dissipation rate of this meta-stable field can have a non-trivial time dependence. We study the impact of this time-dependent dissipation rate on the thermalization of the high-energy decay products of the meta-stable field. These energetic particles can contribute substantially to dark matter production in addition to the usual production from the thermal bath particles during reheating. We investigate the impact of this generalized dissipation on dark matter production in a model-independent way. We illustrate the parameter space that explains the observed dark matter relic abundance in various cosmological scenarios. We observed that dark matter having a mass larger than the maximum temperature attained by the thermal bath can be produced from the collision of the high-energy particles which are not yet thermalized.}


\maketitle
\flushbottom

\section{Introduction}
In standard cosmological theories existence of a meta-stable field dominating the energy density of the universe at the end of inflation is postulated. This meta-stable field can be e.g., an inflaton field \citep{Allahverdi:2002nb, Allahverdi:2010xz}, a curvaton \citep{Moroi:2002rd}, a dilaton \cite{Lahanas:2011tk}, or string theory inspired modulus field \cite{Moroi:1999zb, Starobinsky:1994bd, Dine:1995uk, Acharya:2008bk, Kane:2015jia} etc. The meta-stable field dissipates its energy and produces a thermal bath of relativistic particles or radiation. The universe reheats into a temperature $T_{\text{Reh}}$ via this dissipation and enters into a radiation dominated era at the end of reheating.

In recent years, there have been works \citep{Chung:1998rq, Giudice:2000ex, Allahverdi:2002pu, Allahverdi:2002nb, Allahverdi:2002ap, Kane:2015qea, Garcia:2017tuj, Drees:2017iod, Drees:2018dsj, Allahverdi:2018aux, Arias:2022qjt, Bernal:2022wck, Bhattiprolu:2022sdd} which envisages the possibility of a stable relic production during this matter dominated era after the end of inflation. Ref.~\citep{Giudice:2000ex} has shown that during reheating the thermal bath can attain a significantly higher temperature than $T_{\text{Reh}}$, the temperature at the onset of radiation dominated universe. It has also shown that after reaching the maximum temperature $T_{\text{max}}$, the temperature of the thermal bath falls off differently ($T \propto a^{-3/8}$) in comparison with the usual radiation dominated epoch ($T \propto a^{-1}$). Ref.~\citep{Giudice:2000ex} has also shown that how the relic abundance of dark matter depends on $T_{\text{max}}$ and there can be situations in which significant amount of dark matter can be produced at $T \sim T_{\text{max}}$ \citep{Garcia:2017tuj, Chowdhury:2018tzw, Banerjee:2019asa, Garcia:2020eof, Garcia:2020wiy}. 

These recent works assume one crucial assumption that the radiation created from the decay of the meta-stable scalar field instantaneously thermalizes among themselves and creates a dominating thermal bath of temperature $T$. Recently the validity of this assumption has been questioned. There have been a few works \cite{Garcia:2018wtq, Harigaya:2014waa, Harigaya:2019tzu, Drees:2021lbm, Drees:2022vvn, Mukaida:2022bbo} which relaxes this assumption and tries to address the finiteness of the timescale for thermalization. Taking this into account ref.~\cite{Garcia:2018wtq, Harigaya:2014waa, Harigaya:2019tzu, Drees:2021lbm} have shown that UV-sensitivity of the scattering cross-section significantly enhances dark matter production both before and after thermalization. The timescale for thermalization is vital for determining the abundance of dark matter produced during reheating.     

In the very early stages of reheating, these decay products are highly energetic and less populated compared to their thermal counterpart. These high-energy particles loose their energy and form a thermal distribution of particles having energies much smaller than their progenitors. The thermalization of these high-energy particles, produced via dissipation of the meta-stable field, proceeds via $2 \rightarrow N$ scattering with $N \geq 2$ \citep{Davidson:2000er, Allahverdi:2002pu, Harigaya:2014waa, Mukaida:2015ria, Harigaya:2019tzu, Drees:2022vvn}. $2 \rightarrow2 $ scattering hardly plays any role in attaining a chemical equilibrium, but only redistribute the energy among these particles.  It has been shown in ref.~\citep{Davidson:2000er, Allahverdi:2002pu, Harigaya:2014waa, Harigaya:2019tzu} that 2 $\rightarrow$ 3 scattering process is more efficient for the thermalization process in comparison with 2 $\rightarrow$ 2 scattering. This 2 $\rightarrow$ 3 process can be realized as a 2 $\rightarrow$ 2 process followed by a 1 $\rightarrow$ 2 process, in which any one of the final-state particle splits nearly-collinear in momentum into two low-energy daughter particles. Due to this collinearity, these daughter particles experience multiple scattering before their formation which suppresses the splitting rate \cite{Kurkela:2011ti, Harigaya:2013vwa}. This destructive interference effect is known as Landau-Pomeranchuk-Migdal(LPM) effect. This suppressed splitting rate delays the thermalization, but happens at a time-scale smaller than the time-scale associated with a $2 \rightarrow 2$ scattering process.

It was noted in ~\citep{Mukaida:2012qn,Co:2020xaf} that the dissipation rate of a meta-stable field oscillating in a thermal background depends on the field and its mass or the background temperature. If the background temperature is smaller than the mass of the field, then the dissipation rate only depends on the field and its mass. For a quadratic potential, the amplitude of the field redshifts as $a^{-3/2}$. Thus, the evolution of the field induces a time dependence in the dissipation rate of the field \citep{Co:2020xaf}. Motivated by these scenarios, in this work, we focus on the effect of having a generalized dissipation rate ~\citep{Ai:2021gtg, Wang:2022mvv} on the thermalization of high-energy particles. Presence of such generalized dissipation rate alters the temperature evolution of the universe during reheating as compared to the scenario of a constant dissipation rate. Thus, we study the impact of such generalized dissipation rate on dark matter production via the freeze-in mechanism \citep{Hall:2009bx, Hall:2010jx, Cheung:2011nn, Mambrini:2013iaa, Mambrini:2015vna, Mambrini:2016dca, Dudas:2017kfz, Garny:2017kha, Benakli:2017whb, Dudas:2017rpa, Chowdhury:2018tzw, Bhattacharyya:2018evo, Bernal:2018qlk, Dudas:2018npp, Banerjee:2019asa, Barman:2022tzk}, while considering non-instantaneous thermalization, in a model independent way. We delineate the parameter space satisfying observed dark matter relic for a few specific models. We find that dark matter can still be produced from the scattering of non-thermal high-energy particles in the region where production from the thermalized bath particles is kinematically forbidden. 

This paper is organized as follows. In Sec.~\ref{reheating}, we discuss how the reheating dynamics proceeds through time-dependent dissipation of a meta-stable field. In Sec.~\ref{DM_Prod}, we compute the abundance of dark matter produced during reheating in the presence of a generalized dissipation rate. In Sec.~\ref{DM_rad}, we calculate the total dark matter relic density before and after the onset of radiation dominated era considering non-instantaneous thermalization. We study the viable parameter space for specific models in Sec.~\ref{par_space}, and compare various contributions that contributes to the dark matter relic in Sec.~\ref{Comp_con}. Finally, we summarize our results in Sec.~\ref{sec:conclusion}.

\section{Reheating via a time-dependent dissipation rate of a meta-stable field}
\label{reheating}

During inflation, the universe goes through a period of accelerated exponential expansion. After the end of inflation, a heavy meta-stable field $\phi$ coherently oscillates around the minimum of its potential and then decays into relativistic particles. The meta-stable field dissipates its energy density into its decay products. The dynamics of this energy transfer is described by the following set of Boltzmann equations \citep{Turner:1983he, Ichikawa:2008ne, Garcia:2020wiy}
\begin{eqnarray}
&& \dot{\rho}_{\phi} + 3H(1 + \omega) \rho_{\phi} = -(1 + \omega)\Gamma_{\phi} \rho_{\phi}, \label{BEQ1} \\
&& \dot{\rho}_{\mathrm{R}} + 4H \rho_{\mathrm{R}} = (1 + \omega)\Gamma_{\phi} \rho_{\phi},  \label{BEQ2}
\end{eqnarray}
where $\rho_{\phi}$ is the average energy density of the meta-stable field, $\rho_{\mathrm{R}}$ is the energy density of its decay products,  
$\Gamma_{\phi}$ denotes the dissipation rate of the meta-stable field, $\omega$ is the equation-of-state of the meta-stable field, $H (\equiv \dot{a}/a)$ is the Hubble expansion rate of the universe, where $a$ indicates the scale factor of the Friedmann-Robertson-Walker (FRW) Universe. Using the Friedmann equation, one can express $H$ in terms of $\rho_{\phi}$ and $\rho_{\text{R}}$  and is given by
\begin{eqnarray}\label{Friedmann Eq}
 H^{2} = \cfrac{1}{3M_{\text{pl}}^{2}}\left(\rho_{\phi} + \rho_{\text{R}}\right),
\end{eqnarray}  
where $M_{\mathrm{pl}} = \frac{1}{\sqrt{8\pi G}} = 2.4 \times 10^{18}$ GeV is the reduced Planck mass. 

The conventional assumption is that after inflation, a heavy meta-stable field decays via a time-independent (constant) dissipation rate \citep{Kofman:1997yn, Garcia:2018wtq, Brandenberger:1997yf, Bassett:2005xm, Allahverdi:2000ss, Kofman:1996mv} and reheats the universe.  In principle, the dissipation rate of the meta-stable field $\Gamma_{\phi}$ can depend on the scale factor $a(t)$ of evolution. The dissipation rate of the meta-stable field can evolve with time through its dependence on energy density and field-dependent mass \citep{Shtanov:1994ce, Kofman:1997yn, Ichikawa:2008ne, Co:2020xaf, Garcia:2020wiy, Ahmed:2021fvt, Banerjee:2022fiw}. Depending upon the ratio of $\Gamma_{\phi}/H$, there can be situations in which reheating is incomplete (see ref.~\citep{Ichikawa:2008ne, Co:2020xaf, Garcia:2020wiy}). It was shown in ref.~\citep{Co:2020xaf, Mukaida:2012qn} that the dissipation rate can be expressed as $\Gamma_{\phi} \sim  \frac{m_{\phi}^{r+1}}{M^{k}\bar{\phi}^{r-k}}$, where $m_{\phi}$ is the mass of the decaying field, $\bar{\phi}$ is its amplitude and $M$ is a UV mass-scale associated with the theory. This dissipation rate results from an effective operator obtained through integrating out a heavy particle. The effective operator is valid only when the mass of the integrated particle is larger than the mass of $\phi$, known as the large field limit~\citep{Mukaida:2012qn}. The large field limit can be realized in various cosmological scenarios e.g., in Affleck-Dine baryogenesis~\citep{Affleck:1984fy, Dine:1995kz}, modulated reheating in the presence of a spectator field~\citep{Kofman:2003nx, Fujita:2016vfj, Karam:2020skk}, and axionic dark matter production within the paradigm of kinetic misalignment~\cite{Co:2019jts, Co:2020dya}. Therefore, it is plausible that the universe reheats within the region of validity of the above-mentioned dissipation rate.
Following this, we parametrize the dissipation rate of the meta-stable field as
\begin{eqnarray}\label{eq:3}
\Gamma_{\phi}(a) = \Gamma_{0}\left(\cfrac{a}{a_{\text{end}}}\right)^{k_{1}},
\end{eqnarray}
where $\Gamma_{0}$ is the time-independent part of the dissipation rate of the inflaton, and  $a_{\text{end}}$ is the scale factor at the end of inflation. In the very early stage of reheating, we assume that $\Gamma_{\phi}(a)/H(a)$ $\ll$ 1, and the energy density of the meta-stable field dominates the energy density of the universe. 
The solution of the Eq.~\eqref{BEQ1} and Eq.~\eqref{BEQ2} with the initial conditions $\rho_{\phi}(a_{\text{end}}) \equiv \rho_{\phi, \text{end}} = 3M_{\text{pl}}^{2}H_{\text{end}}^{2}$ and $\rho_{R}(a_{\text{end}}) = 0$ reads
\begin{eqnarray}
&& \rho_{\phi}(z) \approx \rho_{\phi,\text{end}}\, z^{-3(1+\omega)} = 3\,M_{\text{pl}}^{2}\,H_{\text{end}}^{2}\, z^{-3(1+\omega)}, \label{rhoPhi} \\ 
&& \rho_{R}(z) = \cfrac{2(1+\omega)}{2k_{1} - 3\omega + 5}\, \cfrac{\Gamma_{0}}{H_{\text{end}}}\, 3\,M_{\text{pl}}^{2}\,H_{\text{end}}^{2}\left( z^{\frac{2k_{1} - 3\omega + 5}{2}} - 1 \right) z^{-4}, \label{rhoR}
\end{eqnarray}
where, $z$ is a dimensionless variable ($ z \equiv$ $a/a_{\text{end}}$)  and $H_{\text{end}}$ is the Hubble parameter at the end of inflation. The particle produced from the decay of the meta-stable field has energy $m_{\phi}/2$ with $m_{\phi}$ being the mass of the meta-stable field. These high-energy decay products do not thermalize instantaneously, rather take a finite amount of time to form a thermal bath.
For complete understanding of the kinematics of the thermalization process, one needs the evolution of the phase-space distribution of these high-energy decay products. The Boltzmann equation for the process $\phi$ $\rightarrow$ $\gamma$ + $\gamma$ to compute the phase space distribution of these decay products is given by \citep{Hannestad:2004px, Garcia:2018wtq, Ballesteros:2020adh}
\begin{eqnarray}\label{transport eqn}
\cfrac{\partial f_{R}}{\partial t} - H p \cfrac{\partial f_{R}}{\partial p} 
 =  \cfrac{4\pi^{2}}{p^{2}}\,(1 + \omega)\, n_{\phi}\, \Gamma_{\phi}\, \delta\left(p-\cfrac{m_{\phi}}{2}\right).
\end{eqnarray}
The general solution to the Eq.~\eqref{transport eqn} is given by (for  a detailed derivation see appendix \ref{Appendix:A})
\begin{eqnarray}\label{fgamma}
f_{R}(p, t) = \cfrac{8\pi^{2}}{m_{\phi}^{3}(t)}\,\cfrac{(2k_{1}+ 3(1 + \omega))}{|3\omega - 1|}\, n_{R}(t)\, \left(\cfrac{m_{\phi}(t)}{2p}\right)^{\frac{(3(1 + \omega) - 24\omega - 2k_{1})}{2(1-3\omega)}}\, \theta\left(\cfrac{m_{\phi}(t)}{2}-p\right),
\end{eqnarray}

where we have used $n_{R} = \frac{6(1+\omega)}{2k_{1} + 3(1+\omega)}n_{\phi}(t)\Gamma_{\phi}(t)t$ in the regime $\Gamma_{\phi}t \ll 1$.\\
For $\omega = 0$, the above expression reduces to
\begin{eqnarray}\label{fgamma0}
f_{R}(p, t) &=& \cfrac{8\pi^{2}}{m_{\phi}^{3}}\,(2k_{1} + 3)\, n_{R}(t)\, \left(\cfrac{m_{\phi}}{2p}\right)^{\frac{(3 - 2k_{1})}{2}}\, \theta\left(\cfrac{m_{\phi}}{2}-p\right), \nonumber \\
\end{eqnarray}
with the condition $ \frac{a(t)}{a(\hat{t})} = \frac{m_{\phi}}{2p}$, and $\hat{t} = t \left(\frac{2p}{m_{\phi}}\right)^{\frac{3(1+\omega)}{2}}$, where $\hat{t}$ is the cosmic time when $p$ is equal to $m_{\phi}/2$. $n_{\phi}(t)$ is the number density of inflaton that has not decayed upto that instant and $n_{R}(t)$ is the number density of the decay products. The expression in Eq.~\eqref{fgamma0} matches with the phase-space distribution of ref.~\citep{Garcia:2018wtq} in the $k_{1} \to 0$ limit. In the next section we will discuss the thermalization of the decay products of the meta-stable field.

\subsection{Thermalization of the decay products}
We briefly discuss the thermalization of the highly energetic decay products of the meta-stable field in this section. After the end of inflation, the high-energy particles from the decay of the meta-stable field lose their energy either via large-angle scattering or through soft splittings.
Ref.~\citep{Harigaya:2013vwa} has shown that large-angle scattering (scatterings with large momentum exchange) alone is not sufficient to thermalize the decay products of the meta-stable field. Following ref.~\citep{Kurkela:2011ti}, we will consider the thermalization of these high-energy decay products via splitting into low-energy particles mediated by a gauge interaction of coupling strength $\alpha$. During thermalization, the high-energy particles with momenta $p \sim m_{\phi}$ dominate the number and energy densities of the decay products. Since, the mass of the meta-stable field $m_{\phi}(z) = m_{\phi}(a_{\text{end}})z^{-3\omega}$ redshifts faster for nonzero $\omega$, the decays products thermalize faster compared to $\omega = 0$. So, from this section onwards, we only consider the case with $\omega = 0$.  We consider the case having a mono-energetic distribution function of the decay products
\begin{eqnarray}\label{pdnh}
f_{R}(p) \simeq
\begin{cases}
\cfrac{n_{R}}{m_{\phi}^{3}} & \text{for}\; p \sim m_{\phi}, \\
0 & \text{for}\; \text{others},
\end{cases}
\end{eqnarray}
where $n_{R}$ is the number density of the high-energy particles with momenta $p \sim m_{\phi}$. Following ref.~\citep{Kurkela:2011ti}, we express the distribution function in terms of the gauge coupling strength $\alpha = (g^{2}/(4\pi))$ as
\begin{eqnarray}\label{pdalp}
f_{R}(p) \sim
\begin{cases}
\alpha^{-c} & \text{for}\; p \lesssim m_{\phi}, \\
\left(\cfrac{m_{\phi}}{p}\right)^{4} & \text{for}\; p > m_{\phi}.
\end{cases}
\end{eqnarray}
From Eq.~\eqref{pdnh} - \eqref{pdalp} we can write
\begin{eqnarray}
n_{R} = \alpha^{-c}\, m_{\phi}^{3},
\end{eqnarray}
where $c$ is an arbitrary constant. For the same energy density, systems with $c > 0$ is over-occupied and $c < 0$ is under-occupied compared to thermal occupancy.  The number density of these high-energy particles can be written as
\begin{eqnarray}\label{nhard}
n_{R} & \approx & \cfrac{6}{(2k_{1} + 3)}\, n_{\phi}(t)\, \Gamma_{\phi}(t)\,t  \nonumber \\
& = & \cfrac{8}{(2k_{1} + 3)}\left(\cfrac{\Gamma_{\phi} M_{\text{pl}}^{2}}{m_{\phi}^{3}}\right)\, m_{\phi}^{3}\,(m_{\phi} t)^{-1}.
\end{eqnarray}
Thus
\begin{eqnarray}
 \alpha^{-c} = \cfrac{8}{(2k_{1} + 3)}\left(\cfrac{\Gamma_{\phi} M_{\text{pl}}^{2}}{m_{\phi}^{3}}\right)\,(m_{\phi} t)^{-1}.
 \end{eqnarray}
These high-energy decay products dissipate their energy continuously into low-energy particles via multiple collinear splittings. For a collinear splitting (small angle splitting) when the formation of the wave-packet takes a longer period compared to the time-scale associated with elastic scattering ($t_{\text{el}} \sim \Gamma_{\text{el}}^{-1}$), then there will be interference. This out-of-phase interference suppresses the splitting rate between the daughter and the mother particle, which is known as Landau-Pomeranchuk-Migdal(LPM) effect \citep{Landau:1953um, Landau:1953gr, Migdal:1955nv, Migdal:1956tc}. The LPM suppressed splitting rate reads \citep{Kurkela:2011ti} as
\begin{eqnarray}\label{LPM}
 \Gamma_{\text{split}}^{\text{LPM}}(k) \sim \alpha\, t_{\text{form}}^{-1}(k),
\end{eqnarray} 
where $t_{\text{form}}$ is the formation time of a collinear splitting process given by
\begin{eqnarray}\label{tform}
t_{\text{form}}(k) \sim \sqrt{\cfrac{k}{\hat{q}_\text{elastic}}},
\end{eqnarray} 
where $\hat{q}_\text{elastic}$ is mean-squared momentum transfer per unit time.
These low-energy particles produce a thermal bath having an effective temperature $T_{s}$.  
The temperature of the soft-thermal bath alone is sufficient to describe its dynamics. Momentum transfer in this process can be written in terms of $T_{s}$ as
\begin{eqnarray}\label{qelas}
\hat{q}_\text{elastic} \sim \alpha^{2}T_{s}^{3}.
\end{eqnarray}  
At this moment, the low-energy particles dominate the number density, whereas the high-energy particles dominate the energy density. We assume daughter particles with momentum $k_{\text{split}}$ can transfer their energy entirely to the soft-thermal bath. One can get an estimate about the scale $k_{\text{split}}$ by demanding that in time $t$, each high-energy particle produces $\mathcal{O}(1)$ such daughter particles i.e.,
 \begin{eqnarray}\label{LPM Rate}
 \Gamma_{\text{split}}^{\text{LPM}}(k_{\text{split}})\, t \sim 1.
 \end{eqnarray}
 Using Eq.~\eqref{LPM} - \eqref{qelas} we can express $k_{\text{split}}$ as
 \begin{eqnarray}\label{Ksplit}
 k_{\text{split}} \sim \alpha^{4}\,T_{s}^{3}\,t^{2}.
 \end{eqnarray} 
Particles with momentum $k_{\text{split}}$ can split into daughter particles having smaller momentum through a very high splitting rate. 
These particles with momentum $k_{\text{split}}$ transfer all their energy into the soft-thermal bath through multiple splittings in a period much shorter than the formation time of the initial daughter particle. The energy density of the soft-thermal bath is
 \begin{eqnarray}\label{EDsoft}
 \mathcal{E}_{s} \sim T_{s}^{4} \sim n_{R}\, k_{\text{split}}.
 \end{eqnarray}
Solving Eq.~\eqref{LPM Rate} we get the expression for the temperature of the soft-thermal bath and $k_{\text{split}}$
 \begin{eqnarray}\label{softtemp}
 && T_{s} \sim \alpha^{4}\,n_{R}\,t^{2} \sim \alpha^{4-c}\, m_{\phi}\, (m_{\phi}t)^{2}, \\
 && k_{\text{split}} \sim \alpha^{16-3c}\, m_{\phi}\, (m_{\phi}t)^{8}.
 \end{eqnarray}
 These high-energy particles lose their energy completely when the splitting momentum becomes comparable to their maximum momentum, i.e., $k_{\text{split}}|_{z=z_{\text{th}}} \sim m_{\phi}$. At the time $t_{\text{th}}$, the thermalization of the decay products concluded. The corresponding $z$ is given by
\begin{eqnarray}\label{zth}
z_{\text{th}} \simeq \left( \cfrac{2k_{1} + 3}{8} \right)^{\frac{2}{2k_{1} + 5}}\, \alpha^{-\frac{32}{6k_{1} + 15}}\, \left(\cfrac{\Gamma_{0}M_{\text{pl}}^{2}}{m_{\phi}^{3}}\right)^{-\frac{2}{2k_{1} + 5}} \left( \frac{2}{3}\cfrac{m_{\phi}}{H_{\text{end}}}\right)^{-\frac{10}{6k_{1} + 15}}.
\end{eqnarray}
\begin{table}[h]
\caption{ \text{Typical time-scale for thermalization in various scenarios}}
\centering
\begin{tabular}{||l||c | c | c | c | c |c ||}
\hline
\diagbox[width=7.1em]{\makebox[4em]{$k_{1}$}}{$\Gamma_{0}\, (\text{GeV})$}& \multicolumn{2}{c|}{ \makebox[6em]{$10^{-5}$}} & \multicolumn{2}{c|}{ \makebox[5em]{$10^{-10}$}} & \multicolumn{2}{c||}{ \makebox[5em]{$10^{-15}$}}  \\ [0.5ex]  \hline
& \makebox[3em]{$z_{\text{th}}$} & \makebox[3em]{$z_{\text{Reh}}/z_{\text{th}}$} & \makebox[3em]{$z_{\text{th}}$} & \makebox[3em]{$z_{\text{Reh}}/z_{\text{th}}$} & \makebox[3em]{$z_{\text{th}}$} & \makebox[3em]{$z_{\text{Reh}}/z_{\text{th}}$}  \\ \hline\hline
\makebox[4em]{0} & \makebox[3em]{$\mathcal{O}(10^{5})$} & \makebox[3em]{$\mathcal{O}(10^{6})$} & \makebox[3em]{$\mathcal{O}(10^{7})$} & \makebox[3em]{$\mathcal{O}(10^{7})$} & \makebox[3em]{$\mathcal{O}(10^{9})$} & \makebox[3em]{$\mathcal{O}(10^{8})$} \\ \hline
\makebox[4em]{3/2} & \makebox[3em]{$\mathcal{O}(10^{3})$} & \makebox[3em]{$\mathcal{O}(10^{2})$} & \makebox[3em]{$\mathcal{O}(10^{5})$} & \makebox[3em]{$\mathcal{O}(10^{2})$} & \makebox[3em]{$\mathcal{O}(10^{6})$} & \makebox[3em]{$\mathcal{O}(10^{3})$} \\ \hline
\makebox[4em]{3} & \makebox[3em]{$\mathcal{O}(10^{2})$} & \makebox[3em]{$\mathcal{O}(10^{1})$} & \makebox[3em]{$\mathcal{O}(10^{3})$} & \makebox[3em]{$\mathcal{O}(10^{1})$} & \makebox[3em]{$\mathcal{O}(10^{4})$} & \makebox[3em]{$\mathcal{O}(10^{1})$} \\ \hline\hline
\end{tabular}
\label{table:4}
\end{table}

In the above table, we compare the typical time-scale for the thermalization of high-energy decay products. We have set $\alpha = 0.03, m_{\phi} = 3 \times 10^{13}\, \text{GeV}, H_{\text{end}} = 10^{12}\, \text{GeV}$ to obtain $z_{\text{th}}$ for different $k_{1}$. We see that the time-scale for thermalization differs by orders of magnitude in various cosmological scenarios. The time needed for thermalization decreases with an increasing dissipation rate.\\
Using the expression of $t_{\text{th}}(z_{\text{th}})$ in Eq.~\eqref{softtemp}, we get the thermalization temperature $T_{\text{th}}$ as 
\begin{eqnarray}\label{Tth}
T_{\text{th}} \sim \left( \cfrac{8}{2k_{1} + 3} \right)^{\frac{2}{2k_{1} + 5}}\, \alpha^{\frac{12 - 8k_{1}}{6k_{1} + 15}}\,m_{\phi}\, \left(\cfrac{\Gamma_{0}M_{\text{pl}}^{2}}{m_{\phi}^{3}} \right)^{\frac{2}{2k_{1} + 5}} \left( \cfrac{2}{3} \frac{m_{\phi}}{H_{\text{end}}}\right)^{-\frac{4k_{1}}{6k_{1} + 15}}.
\end{eqnarray}
This generalized expression of $T_{\text{th}}$ matches with the expression of \cite{Garcia:2018wtq} in $k_{1} \to 0$ limit.
\subsection{Spectrum of high-energy decay products after thermalization}
The decay products are thermalized at $z_{\text{th}}$, corresponding to a temperature $T_{\text{th}}$.  The thermalized sector dominates the energy density, as well as the number density of radiation after the thermalization, i.e., for $z$ $>$ $z_{\text{th}}$.
The meta-stable field decay does not end at $z_{\text{th}}$. It rather continuously supplies high-energy particles with energy $E \simeq m_{\phi}/2 > T (z > z_{\text{th}})$ (temperature of the particles in the thermal bath). These high-energy particles take a finite time to thermalize.
We will now estimate the spectrum of these particles with the background thermal bath in the next section.

The phase-space distribution of these out-of-equilibrium particles depends on the magnitude of their three-momentum $p$ and time $t$. In a collinear splitting process, the transverse momenta of the daughter particles are negligible. We consider the phase space distribution function in one dimension\footnote{One can write the number density as $n_{R}(t) = \int \tilde{f}_{R}(p, t)\, dp$.}. The relevant Boltzmann equation governing the splitting processes is \citep{Harigaya:2014waa, Drees:2021lbm}
\begin{eqnarray}\label{eq:3.4}
\cfrac{\partial \tilde{f}_{R}}{\partial t} - 3 H p \cfrac{\partial \tilde{f}_{R}}{\partial p} &=& C[\tilde{f}_{R}(p,t)]\nonumber \\
& = & 2\, n_{\phi} \Gamma_{\phi} \delta\left(p - \cfrac{m_{\phi}}{2}\right) + \int_{2p}^{m_{\phi}/2} dk \cfrac{d\Gamma^{\text{split}}_{\text{LPM}}}{dp}(p)\tilde{f}_{R}(k,t) \nonumber \\
&& + \int_{T}^{p} dk\cfrac{d\Gamma^{\text{split}}_{\text{LPM}}}{dk}(k)\tilde{f}_{R}(p+k,t) -
\int_{T}^{p/2} dk\cfrac{d\Gamma^{\text{split}}_{\text{LPM}}}{dk}(k)\tilde{f}_{R}(p,t). \nonumber\\
\end{eqnarray}
The first term on the right-hand side of Eq.~\eqref{eq:3.4} represents the contribution from the decay of the meta-stable field. The second and third terms describe the injection of particles with momentum $p$ from the splitting of higher momentum ($>p$) particles. The last term illustrates the depletion of particles with a momentum of $p$ due to splitting into daughter particles of momentum $k$. We use a hard infrared cutoff to regularize the infrared divergences in the last two integrals. This will not affect the result as long as $p \gg T$. We avoid the collinear divergences by demanding a minimum splitting angle $\theta_{\text{min}}$ which is of $\mathcal{O}(\left(T/m_{\phi}\right)^{3/4})$.
The splitting rate of high-energy decay products gets suppressed due to the LPM effect. The LPM suppressed splitting rate can be written as \cite{Kurkela:2011ti, Drees:2021lbm}
\begin{eqnarray}\label{eq:3.5}
\cfrac{d\Gamma^{\text{split}}_{\text{LPM}}}{dk} \simeq \alpha^{2}\left(\cfrac{T}{k}\right)^{\frac{3}{2}} \sqrt{\tilde{g_{*}}},
\end{eqnarray}
where $\tilde{g_{*}}$ denotes the number of degrees of freedom of these low-energy particles in the thermal bath.
Solving Eq.~\eqref{eq:3.4} using Eq.~\eqref{eq:3.5} we get \cite{Harigaya:2014waa, Drees:2021lbm}
\begin{eqnarray}\label{ftilde}
\tilde{f}_{R}(p) \simeq \cfrac{n_{\phi}\,\Gamma_{\phi}}{\sqrt{\tilde{g_{*}}}\, \alpha^{2}\, T^{3/2}}\,m_{\phi}\,p^{-3/2}.
\end{eqnarray}

The decay products of the meta-stable field are thermalized at $z_{\text{th}}$. The thermalized sector dominates the energy density of radiation after thermalization. The temperature of the thermal bath after thermalization evolved as
\begin{eqnarray}\label{eq:Temp}
T(z) = \left(\cfrac{30}{\pi^{2}g_{*}(T)} \right)^{1/4}\left[\cfrac{2}{2k_{1} + 5}\, \cfrac{\Gamma_{0}}{H_{\text{end}}}\, 3 M_{\text{pl}}^{2}\, H_{\text{end}}^{2} \right]^{1/4}\, z^{\frac{2k_{1} - 3}{8}},
\end{eqnarray}
where $g_{*}(T)$ is the number of relativistic degrees of freedom of the thermal bath at temperature $T$.
From the Eq.~\eqref{eq:Temp}, one can see that depending upon the value of $k_{1}$, the evolution of temperature during reheating can be divided into three categories:
\begin{table}[h]
\caption{ \text{Temperature evolution during reheating}}
\centering
\begin{tabular}{||c | c| c| c | c  ||} 
 \hline
 & Condition & Temperature & $T_{\text{min}}$ & $T_{\text{max}}$ \\ [0.5ex] 
\hline\hline
\text{Case A} & $2k_{1} > 3$ & increases        & $T_{\text{th}}$ & $T_{\text{Reh}}$ \\ \hline
\text{Case B} & $2k_{1} = 3$ & constant & $T_{\text{Reh}}$ & $T_{\text{Reh}}$ \\ \hline
\text{Case C} & $2k_{1} < 3$ & decreases        & $T_{\text{Reh}}$ & $T_{\text{th}}$ \\
 \hline\hline
\end{tabular}
\end{table}

Reheating concludes when the energy density of the meta-stable field and radiation becomes equal,  i.e., $\rho_{\phi}(z_{\mathrm{Reh}})$ $=$ $\rho_{R}(z_{\mathrm{Reh}})$ and the corresponding $z$ is given by
\begin{eqnarray}
z_{\text{Reh}} = \cfrac{\left(3 M_{\text{pl}}^{2} H_{\text{end}}^{2}\right)^{\frac{2}{2k_{1} + 3}}}{\left(\cfrac{2}{2k_{1} + 5}\, \cfrac{\Gamma_{0}}{H_{\text{end}}}\, 3 M_{\text{pl}}^{2}\, H_{\text{end}}^{2} \right)^{\frac{2}{2k_{1} + 3}}}.
\end{eqnarray} 
The temperature of the thermal bath corresponding to $z_{\mathrm{Reh}}$ is given as
\begin{eqnarray}\label{Treh}
T_{\text{Reh}} &=& \left( \cfrac{30}{\pi^{2}\, g_{*}(T_{\text{Reh}})} \right)^{1/4}\, \left(\cfrac{2}{2k_{1} + 5}\, \cfrac{\Gamma_{0}}{H_{\mathrm{end}}}\, 3 M_{\text{pl}}^{2}\, H_{\text{end}}^{2}\right)^{1/4}\, z_{\text{Reh}}^{\frac{2k_{1} - 3}{8}} \nonumber \\
& = & \left( \cfrac{30}{\pi^{2}\, g_{*}(T_{\text{Reh}})} \right)^{1/4}\, \cfrac{(3 M_{\text{pl}}^{2}\, H_{\text{end}}^{2})^{1/4}}{z_{\text{Reh}}^{3/4}}.
\end{eqnarray}
The particles in the thermal bath dominate the energy density of the universe after reheating, and the bath temperature redshifts inversely with the scale factor ($a^{-1}$).

\section{Dark matter production during reheating}\label{DM_Prod}
In the previous section, we have derived the phase-space distribution of the non-thermal\footnote{phase-space distribution differs from the particles in a thermal bath.} particles for time-dependent dissipation of the meta-stable field. In this section, we consider the impact of these highly energetic particles on dark matter production through freeze-in during the reheating of the universe. We assume that the dark matter does not couple directly to the meta-stable field. Therefore, we are not considering dark matter production from the meta-stable field. We concentrate on dark matter production from the collision between the high-energy decay products of the meta-stable field and the particles in the thermal bath.
There are four production mechanisms of dark matter during reheating
\[
\mathrm{DM\, Production} \rightarrow
\begin{cases}
\text{Pre-thermal} \hspace*{0.25cm}\rightarrow

\text{Non-thermal(I)} \hspace*{1.55cm} z_{\text{end}} < z < z_{\text{th}}\\
\\
\text{Post-thermal} \rightarrow 
\begin{rcases}
\text{Non-thermal(II)} \\
\text{Non-thermal(III)} \\
\text{Thermal(IV)}
\end{rcases}
\hspace*{0.75cm} z_{\text{th}} < z < z_{\text{Reh}}
\end{cases}
\]
\begin{itemize}
\item \text{Non-thermal(I)}: Production of dark matter from high-energy particles before thermalization.
\item \text{Non-thermal(II)}: Dark matter production from the collision between two high-energy particles after thermalization.
\item \text{Non-thermal(III)}: Production of dark matter from the collision between a high-energy particle and a particle in the thermal bath.
\item \text{Thermal(IV)}: Dark matter production from the collision between two particles in thermal bath.
\end{itemize}
The total cross-section for dark matter production can be parametrized as \citep{Garcia:2018wtq, Harigaya:2019tzu}
\begin{align}
\label{xs_param}
\sigma (s) = \lambda c_{n} \cfrac{s^{n/2}}{M^{n+2}}\; ,
\end{align}
where $\sqrt{s}$ is the center of mass energy of the process, $M$ is a heavy mass-scale relevant to the process, $\lambda$ is a dimensionless coupling and $c_{n} = \frac{32}{2^{4+n}}\, \frac{\zeta(3)^{2}}{\pi} \frac{1}{\Gamma\left(2 + \frac{n}{2}\right)\, \Gamma\left(3 + \frac{n}{2}\right)}$. While parametrizing the cross section as expressed in Eq.~\eqref{xs_param}, we have assumed $m_{\chi}, m_{\mathrm{inc}}, \sqrt{s} \ll M$, with $m_{\mathrm{inc}}$ being the mass of the incoming particles. Cross-section with $n>0$ can be due to the interaction of dark matter with visible sector particles via a mediator having derivative couplings \citep{Mambrini:2013iaa,Mambrini:2015vna,Soni:2016gzf, Benakli:2017whb,Dudas:2017rpa,Chowdhury:2018tzw}, where $M$ can be identified as the cutoff scale of the theory.
\subsection{DM production before thermalization}
Since we assume that dark matter does not yield from the decay of the meta-stable field, the only production channel available for dark matter production before thermalization is $\gamma(k_{1})$ $+$ $\gamma(k_{2})$ $\rightarrow$ $\chi(p_{1})$ $+$ $\chi(p_{2})$. In this section, we consider dark matter production before thermalization from these high-energy decay products. The Boltzmann equation for the dark matter number density can be written as
\begin{eqnarray}\label{BEQbnth}
\dot{n_{\chi}} + 3 H n_{\chi} &=& 4\, g_{\chi}^{2}\, g_{R}^{2}\, \int \cfrac{d^{3}k_{1}}{(2\pi)^{3}2k_{1}^{0}}\, \int \cfrac{d^{3}k_{2}}{(2\pi)^{3}2k_{2}^{0}}\, \left(k_{1}\cdot k_{2}\right)\, \sigma(s)\, f_{R}(k_{1})\, f_{R}(k_{2}) \nonumber \\
& \equiv & \langle \sigma v \rangle _{NT}\, n_{R}^{2},
\end{eqnarray}
where $\sigma(s)$ is the scattering cross-section for the process $\gamma + \gamma \rightarrow \chi  + \chi$.  $g_{\chi}$ and $g_{R}$ denotes the number of degrees of freedom (d.o.f) of $\chi$ and $\gamma$, respectively. Using the phase-space distribution function of the high-energy decay products given in Eq.~\eqref{fgamma0}, we get $\langle \sigma v \rangle _{NT}$ as
\begin{eqnarray}\label{sigmav}
\langle \sigma v \rangle _{NT} = \, g_{\chi}^{2}\, g_{R}^{2}\, \cfrac{(2k_{1} + 3)^{2}}{m_{\phi}^{(3 + 2k_{1})}} \int _{0}^{m_{\phi}^{2}} ds\, \sigma(s)\,s\,\int _{\sqrt{s}}^{m_{\phi}}dE_{+}  \int _{-\sqrt{E_{+}^{2}-s}}^{\sqrt{E_{+}^{2}-s}}dE_{-}\, \cfrac{1}{\left(E_{+}^{2}-E_{-}^{2}\right)^{\frac{(3 - 2k_{1})}{2}}}. \nonumber \\
\end{eqnarray}
The comoving number density of dark matter can be defined as $Y_{\chi}(z) \equiv n_{\chi}(z)\, z^{3}$. We rewrite Eq.~\eqref{BEQbnth} in terms $Y_{\chi}(z)$. We get
\begin{eqnarray}\label{BEQCNBTh}
\cfrac{d Y_{\chi}}{d z} = \cfrac{\langle \sigma v \rangle _{NT}\, n_{R}^{2}\, z^{3}}{z H}.
\end{eqnarray}
As, the meta-stable field dominates the energy density of the universe during reheating, Hubble parameter $H(z)$ can be expressed as
\begin{eqnarray}\label{eq:hubble}
H(z) = H_{\text{end}}\, z^{-\frac{3}{2}}.
\end{eqnarray}
Solving Eq.~\eqref{BEQCNBTh}, we get the amount of  dark matter produced from the high-energy decay products until thermalization
\begin{eqnarray}\label{chint}
Y_{\chi}^{(NT)}(z_{\text{th}}) =\cfrac{2}{(4k_{1} + 3)} \cfrac{\langle \sigma v \rangle_{\text{NT}}}{H_{\text{end}}}\, \left(\cfrac{3M_{\text{pl}}^{2}H_{\text{end}}^{2}}{m_{\phi}}\right)^{2}\, \frac{z_{\text{th}}^{\left(\frac{4k_{1} + 3}{2}\right)} - 1}{z_{\text{Reh}}^{(2k_{1} + 3)}}.
\end{eqnarray}
\subsection{Dark matter production after thermalization}
For $z < z_{\text{th}}$, the entire contribution in dark matter number density comes from the high-energy particles. Whereas, for $z > z_{\text{th}}$, there exists a thermal bath of particles with characteristics temperature $T$. We now focus on the dark matter produced during the period $z_{\text{th}}$ $<$ $z$ $<$ $z_{\text{Reh}}$. The relevant Boltzmann equation has the form
\begin{eqnarray}\label{BEQg}
\dot{n}_{\chi} + 3 H n_{\chi}  = R(T(z)),
\end{eqnarray}
where $R(T)$ denotes the production rate of dark matter. We can express the production rate as
\begin{eqnarray}
R(T(z)) \equiv 4\, g_{\chi}^{2}\, g_{R}^{2}\, \int \cfrac{d^{3}k_{1}}{(2\pi)^{3}2k_{1}^{0}}\, \int \cfrac{d^{3}k_{2}}{(2\pi)^{3}2k_{2}^{0}}\, \left(k_{1}\cdot k_{2}\right)\, \sigma(s)\, f_{R}^{\text{psth}}(k_{1})\, f_{R}^{\text{psth}}(k_{2}).\nonumber \\
\end{eqnarray}
Where $f_{R}^{\text{psth}}$ denotes the phase-space distribution of particles after thermalization, it can be either $\tilde{f}_{R}(p)$ given in Eq.~\eqref{ftilde} or $f_{\xi i}^{(\text{eq})}$ where $f_{\xi i}^{(\text{eq})}$ is the equilibrium phase space distribution, and it has the form $f_{\xi i}^{(\text{eq})} = e^{(-E_{i}/T)}$, $E_{i}$ is the energy of $\xi_{i}$.
The Eq.~\eqref{BEQg} can be expressed in terms of the comoving number density of dark matter
and  the parameter $z$ as
\begin{eqnarray}\label{eq:3.10}
\cfrac{d Y_{\chi}}{d z} =  \cfrac{R(T(z))z^{3}}{zH}.
\end{eqnarray}
The solution of the Eq.~\eqref{eq:3.10} gives the comoving number density of dark matter produced during the period $z_{\text{th}}$ $<$ $z$ $<$ $z_{\text{Reh}}$ and is given by
\begin{eqnarray}
Y_{\chi}(z_{\text{Reh}}) = Y_{\chi}^{(\text{NT})}(z_{\text{th}}) + \int_{z_{\text{th}}}^{z_{\text{Reh}}}\, dz\, \cfrac{R(T(z))z^{3}}{zH},
\end{eqnarray}
where $Y_{\chi}^{(\text{NT})}(z_{\text{th}})$ is given in Eq.~\eqref{chint}.
The task left for us is to calculate the production rate of dark matter for various production processes in this regime. As discussed earlier, there are three processes that contribute in dark matter production during the period $z_{\text{th}}$ $<$ $z$ $<$ $z_{\text{Reh}}$ and is given by
\begin{itemize}
\item collision between two high-energy particles.
\item collision between a high-energy particle and a particle in thermal bath.
\item a collision between two particles in thermal bath.
\end{itemize} 
\subsubsection*{Non-thermal(II): Collision between two non-thermal particle with E$_{1}$, E$_{2}$ > T}
High-energy decay products are continuously produced from the decay of the meta-stable field until reheating. They take a finite time to attain a equilibrium with the thermal bath. Before their thermalization, these high-energy particles can collide with each other and yield a pair of dark matter particles if the production is kinematically available i.e., $m_{\chi} < m_{\phi}/2$.
In this subsection, we calculate the number density of dark matter produced from the collision between two high energy particles with $T$ $<$ $E_{1}, E_{2}$ $\leq$ $m_{\phi}/2$. This production channel is kinematically allowed as long as $E_{1} E_{2} \geq m_{\chi}^{2}$. We designate this production process as \text{Non-thermal(II)}.
The production rate of dark matter for the collision between two high-energy particles can be expressed as
\begin{eqnarray}
R_{(p)}^{(\text{NT})}(T(z)) & = &  \cfrac{2^{n+1}}{(n-1)^{2}} \cfrac{g_{\chi}^{2}\,g_{R}^{2}}{\pi^{2}}\cfrac{\lambda c_{n}}{M^{n+2}} \cfrac{9\, M_{\text{pl}}^{4}\, H_{\text{end}}^{4}\, \Gamma_{0}^{2}}{\tilde{g_{*}}\, \alpha^{4}}\, \left(E_{1}E_{2}\right)^{\frac{n-1}{2}}\, z^{(2k_{1} - 6)} \nonumber \\
& = & \cfrac{2^{n+1}}{(n-1)^{2}} \cfrac{g_{\chi}^{2}\,g_{R}^{2}}{\pi^{2}}\cfrac{\lambda c_{n}}{M^{n+2}} \cfrac{9\, M_{\text{pl}}^{4}\, H_{\text{end}}^{4}\, \Gamma_{0}^{2}}{\tilde{g_{*}}\, \alpha^{4}}\, \left( \left(\cfrac{m_{\phi}}{2}\right)^{\frac{n-1}{2}} - m_{\chi}^{\frac{n-1}{2}} \right)^{2} z^{(2k_{1} - 6)}. \nonumber \\ 
\end{eqnarray}
\subsubsection*{Non-thermal(III): Collision between high energy particle with E$_{1}$ > T and a particle in the thermal bath}
Until now, we only considered dark matter production from non-thermal particles. We did not take into account the contribution of the thermal bath particles to dark matter production. But the collision of a high-energy particle and a particle in the thermal bath can produce a substantial amount of dark matter.
In this subsection, we focus on calculating the amount of dark matter produced from the collision between a high energy particle with energy $T$ $<$ $E_{1}$ $\leq$ $\frac{m_{\phi}}{2}$ and a particle in the thermal bath with characteristic temperature $T$. This production channel is kinematically available as long as $E_{1} T \gtrsim m_{\chi}^{2}$. We label this production channel as \text{Non-thermal(III)}. The production rate of dark matter for this channel can be written as
\begin{eqnarray}
R_{(m)}^{(\text{NT})}(T(z)) & = & \cfrac{2^{n}}{(n-1)} \cfrac{g_{\chi}^{2} g_{R}^{2}}{\pi^{2}} \cfrac{\zeta(3)}{\pi^{2}} \cfrac{\lambda c_{n}}{M^{n+2}} \cfrac{3\, M_{\text{pl}}^{2}\, H_{\text{end}}^{2}\, \Gamma_{0}}{\sqrt{\tilde{g_{*}}}\, \alpha^{2}} (E_{1})^{(n-1)/2}\, z^{(k_{1} - 3)}\, T^{(n+3)/2} \nonumber\\
& = & \cfrac{2^{n}}{(n-1)}\, \cfrac{g_{\chi}^{2} g_{R}^{2}}{\pi^{2}}\, \cfrac{\zeta(3)}{\pi^{2}} \cfrac{\lambda c_{n}}{M^{n+2}}\, \cfrac{3\, M_{\text{pl}}^{2}\, H_{\mathrm{end}}^{2}\, \Gamma_{0}}{\sqrt{\tilde{g_{*}}}\, \alpha^{2}}\, \left( \left(\cfrac{m_{\phi}}{2}\right)^{\frac{n-1}{2}} - \left(E_{1}^{\text{min}}\right)^{\frac{n-1}{2}} \right)\nonumber \\ & & \times z^{(k_{1} - 3)} T^{(\frac{n+3}{2})}.
\end{eqnarray}
One can estimate the minimum value of $E_{1}$ required for the pair production of dark matter and is given by
\begin{eqnarray}
E_{1}^{\text{min}} \approx \cfrac{m_{\chi}^{2}}{T_{\text{min}}},
\end{eqnarray}
where $T_{\text{min}}$ is the minimum temperature achieved by the thermal bath during reheating. $T_{\text{min}}$ can be either $T_{\text{th}}$ or $T_{\text{Reh}}$ depending upon the value of $k_{1}$.
The comoving number density of non-thermal dark matter at a redshift $z$ can be expressed as
\begin{eqnarray}
Y_{\chi}^{(\text{NT})}(z) = Y_{\chi}^{(\text{NT})}(z_{\text{th}})+ \int_{z_{\text{th}}} ^{z} \cfrac{\left(R_{(p)}^{(\text{NT})}(T(z)) + R_{(m)}^{(\text{NT})}(T(z))\right)}{H(z)} \,z^{2}\, dz.
\end{eqnarray}
\subsubsection*{Thermal(IV): Collision between two particles in thermal bath}
After $z > z_{\text{th}}$, there exists a thermal bath of particles. These particles can contribute significantly in dark matter production.
In this subsection, we concentrate on determining the abundance of dark matter produced from the collision between two particles in thermal bath. We denote this process as \text{Thermal(IV)}. The production rate of dark matter in this case is given as
\begin{eqnarray}\label{thrate}
R_{(\text{th})}(T) & = & 2 g_{R}^{2}\int d\Pi_{\chi_{1}}\, d\Pi_{\chi_{2}}\, f_{\chi_{1}}^{(\text{eq})}\, f_{\chi_{2}}^{(\text{eq})}\, \lambda^{1/2}(s, m_{\chi_{1}}, m_{\chi_{2}})\, \sigma_{ \chi \chi \rightarrow \gamma\gamma}(s) \nonumber \\
& = & \cfrac{g_{\chi}^{2}\, g_{R}^{2}}{32 \pi^{4}}\, T\int_{4m_{\chi}^{2}}^{\infty} ds\, (s - 4m_{\chi}^{2})\sqrt{s}\, \sigma_{\chi \chi \rightarrow \gamma\gamma}(s)\, K_{1}[\sqrt{s}/T] 
\nonumber \\ 
& = & \cfrac{g_{\chi}^{2}\, g_{R}^{2}}{32 \pi^{4}}\,\cfrac{\lambda\, c_{n}}{M^{n+2}} \, T\int_{4m_{\chi}^{2}}^{\infty} ds\, (s - 4m_{\chi}^{2})\sqrt{s}\, s^{n/2} \, K_{1}[\sqrt{s}/T], 
\end{eqnarray}
where $f_{\chi i}^{(\text{eq})}$ is the equilibrium phase space distribution and it has the form $f_{\chi i}^{(\text{eq})} = \text{exp}(-E_{i}/T)$, $E_{i}$ is the energy of $\chi_{i}$. The function $\lambda$ is given by
\begin{eqnarray}
\lambda(s, m_{\chi_{1}}, m_{\chi_{2}}) \equiv \left[s - \left(m_{\chi_{1}} + m_{\chi_{2}}  \right)^{2} \right]\left[s - \left(m_{\chi_{1}} - m_{\chi_{2}}  \right)^{2} \right].
\end{eqnarray}
The comoving number density of thermal dark matter at a redshift $z$ can be expressed as
\begin{eqnarray}
Y_{\chi}^{(T)}(z) = Y_{\chi}^{(T)}(z_{\text{th}}) + \int_{z_{\text{th}}} ^{z} dz \cfrac{z^{2}}{H(z)} R_{\text{th}}(T(z)) = \int_{z_{\text{th}}} ^{z} dz \cfrac{z^{2}}{H(z)} R_{\text{th}}(T(z)), \nonumber \\
\end{eqnarray}
where the initial number density of thermal dark matter is $Y_{\chi}^{(T)}(z_{\mathrm{th}}) = 0$.

\section{Dark matter relic density with non-instantaneous thermalization}\label{DM_rad}
In this section, we calculate the total dark matter relic density accumulated since the end of inflation until the present day. We begin this section by discussing the dark matter production after reheating is complete. The solution of the Boltzmann equation (Eq.~\eqref{BEQg}) gives \citep{Garcia:2017tuj, Chowdhury:2018tzw, Banerjee:2022fiw}
\begin{eqnarray}\label{eq:4.1}
Y_{\chi}(z_{0}) = Y_{\chi}(z_{\text{Reh}}) + \cfrac{\sqrt{3}\, M_{\text{pl}}}{\rho_{R}(z_{\text{Reh}})^{1/2}}\, \int_{z_{\text{Reh}}}^{z_{0}}\,  R(T(z))\, z^{4}\, dz,
\end{eqnarray}
where $R(T)$ denotes the production rate of dark matter. It can be written as
\begin{eqnarray}\label{eq:4.2}
R(T(z)) = \cfrac{\lambda}{M^{n+2}}\, \cfrac{\zeta(3)^{2}}{\pi^{5}}\, \left(\cfrac{30}{\pi^{2}\, g_{*}(T)}\right)^{\frac{n+6}{4}}\,  \rho_{R}(z_{\text{Reh}})^{\frac{n+6}{4}}\,z^{-(6+n)}.
\end{eqnarray}
Integrating Eq.~\eqref{eq:4.1} using Eq.~\eqref{eq:4.2} gives,
\begin{eqnarray}
Y_{\chi}(z_{0}) &=& Y_{\chi}(z_{\text{Reh}}) + \sqrt{3} M_{\text{pl}} \cfrac{\lambda}{M^{n+2}}\, \cfrac{\zeta(3)^{2}}{\pi^{5}}\, \left(\cfrac{30}{\pi^{2}\, g_{*}(T)}\right)^{\frac{n+6}{4}}\, \nonumber\\ && \times \rho_{R}(z_{\text{Reh}})^{1 + \frac{n}{4}}\, \cfrac{1}{(1+n)} 
\left(\cfrac{1}{z_{\text{Reh}}^{n+1}} - \cfrac{1}{z_{0}^{n+1}} \right).
\end{eqnarray}
Dark matter production from the thermal bath particles occurs dominantly during reheating for $n > -1$. On the other hand, for $n < -1$ production dominantly happens after reheating. The present-day relic density of dark matter is given by \cite{Kolb:1990vq},
\begin{eqnarray}
\Omega_{\chi}\,h^{2} = \cfrac{m_{\chi}\, n_{\chi}(T_{\mathrm{0}})}{\rho_{\text{c}}}\, = \cfrac{m_{\chi}\, Y_{\chi}(z_{0})}{\rho_{\text{c}}\,z_{0}^{3}},
\end{eqnarray}
where $\rho_{\text{c}}$,\footnote{The numerical value of $\rho_{c}$ based upon present day Hubble expansion rate is given by,\\ $\rho_{c}$ = $3\, M_{\text{pl}}^{2}\, H_{0}^{2}$ = $\left( 2.5\times 10^{-3}\, \right)^{4} \text{eV}^{4}$.} denotes the critical density today and $z_{0}$ is present-day redshift of the universe. Thus, the total relic density of dark matter can be written as the sum of all the contributions 
\begin{eqnarray}
\Omega_{\chi}h^{2} = \Omega_{\chi,\, NT}^{(1)}h^{2} + \Omega_{\chi,\, NT}^{(2)}h^{2} + \Omega_{\chi,\, NT}^{(3)}h^{2} + \Omega_{\chi,\, Th}h^{2}.
\end{eqnarray}
\begin{itemize}
\item $ \Omega_{\chi,\, \text{NT}}^{(1)}\,h^{2}:$  Relic density of dark matter produced from the collision between two high-energy decay products before thermalization.
\item $ \Omega_{\chi,\, \text{NT}}^{(2)}\,h^{2}:$ Relic density of dark matter produced from the collision between two high-energy decay products after thermalization.
\item $ \Omega_{\chi,\, \text{NT}}^{(3)}\,h^{2}:$ Relic density of dark matter produced from the collision between a particle in thermal bath and a high-energy decay product after thermalization.
\item $ \Omega_{\chi,\, \text{Th}}\,h^{2}:$ Relic density of dark matter produced from the collision between two particles in thermal bath after thermalization.
\end{itemize}
The relic density of dark matter produced from the scattering of two 
high-energy particles after thermalization has the expression
\begin{eqnarray}\label{nont2}
\Omega_{\chi,\, \text{NT}}^{(2)} = && g_{\chi}^{2}\, g_{R}^{2}\, \cfrac{2^{(n+1)}}{(n-1)^{2}}\, \cfrac{4}{\pi^{2}}\, \left(\cfrac{2}{10k_{1} - 3}\right)\, \left(\cfrac{2k_{1} + 5}{2} \right)^{2}\, \cfrac{\lambda c_{n}}{g_{*}\alpha^{4}}\nonumber \\
 &&  \times \left(\cfrac{\pi^{2}}{30}\,g_{*}(T)\right)^{3/4}\, \left(\cfrac{\pi^{2}}{30}\,g_{*}(T_{\text{Reh}})\right)\, \left(\cfrac{\pi^{2}}{30}\,g_{*}(T_{0})\right)^{3/4} \nonumber \\
&& \times \left[\cfrac{m_{\chi}}{M^{n+2}}\, T_{\text{Reh}}^{4}\, \cfrac{1}{\sqrt{3}M_{\text{pl}}}\left(\left(\cfrac{m_{\phi}}{2} \right)^{\frac{(n-1)}{2}} - m_{\chi}^{\frac{(n-1)}{2}}\right)^{2}\, \cfrac{T_{0}^{3}}{\rho_{c}} \right],
\end{eqnarray}
for $k_{1} > 3/10$. One can obtain the expression for $k_{1} < 3/10$ by appropriately expressing $z_{\text{th}}(T_{\text{th}})$ in terms of $T_{\text{Reh}}$.\\
Depending upon the value of $n$,
the above expression can be approximated as
\begin{eqnarray}\label{Nt2approx}
\Omega_{\chi,\, \text{NT}}^{(2)} \sim
\begin{cases}
m_{\chi}\, T_{\text{Reh}}^{4} &  \text{for} \;n > 1, \\
m_{\chi}\, T_{\text{Reh}}^{4}\, \left(\left(\cfrac{2}{m_{\phi}}\right)^{(1-n)/2} - \cfrac{1}{m_{\chi}^{(1-n)/2}}\right)^{2}\ & \text{for} \;n < 1.
\end{cases}
\end{eqnarray}
The above expression of relic density for $n < 1$ can be re-expressed as
\begin{eqnarray}
\Omega_{\chi,\, NT}^{(2)} \sim
m_{\chi}\, T_{\text{Reh}}^{4}\, \cfrac{\left(m_{\phi}^{(1-n)/2} - (2m_{\chi})^{(1-n)/2} \right)^{2}}{(m_{\phi}\, m_{\chi})^{(1-n)}}.
\end{eqnarray}
The relic density of DM produced from the collision between a particle in thermal bath and a high-energy particle after thermalization can be expressed as
\begin{eqnarray}\label{nont3}
\Omega_{\chi,\, \text{NT}}^{(3)} = && g_{\chi}^{2}\, g_{R}^{2}\, \cfrac{2^{(n+1)/2}}{(n-1)}\, \cfrac{\zeta(3)}{\pi^{4}}\, \cfrac{\lambda c_{n}}{\sqrt{g_{*}}\alpha^{2}}\, \cfrac{2k_{1} + 5}{2\,C_{1}}\nonumber \\
&& \times \left(\cfrac{30}{\pi^{2}g_{*}(T)}\right)^{(n+3)/8}\, \left(\cfrac{\pi^{2}}{30}\,g_{*}(T_{\text{Reh}})\right)^{(n+5)/8}  \left(\cfrac{\pi^{2}}{30}\,g_{*}(T_{0})\right)^{3/4} \nonumber \\
&& \times  \left[\cfrac{m_{\chi}}{M^{n+2}}\,T_{\text{Reh}}^{(n+5)/2}\left(m_{\phi}^{(n-1)/2} - \left(\cfrac{2m_{\chi}^{2}}{T_{\text{min}}}\right)^{(n-1)/2} \right)\, \cfrac{T_{0}^{3}}{\rho_{c}}   \right], \nonumber \\
\end{eqnarray}
where $C_{1} = \frac{n+11}{8}\,\frac{2k_{1} - 3}{2} + 3$.\\
We can approximate the expression of $\Omega_{\chi,\, \text{NT}}^{(3)}$ depending upon the value of $n$ and the approximate expression is given as
 \begin{eqnarray}
\Omega_{\chi,\, \text{NT}}^{(3)} \sim
\begin{cases}
m_{\chi}\, T_{\text{Reh}}^{(n+5)/2} &  \text{for} \;n > 1, \\
m_{\chi}\, T_{\text{Reh}}^{(n+5)/2}\, \left(\left(\cfrac{T_{\text{min}}}{2m_{\chi}^{2}}\right)^{(1-n)/2} - \cfrac{1}{m_{\phi}^{(1-n)/2}}\right)\ & \text{for} \;n < 1.
\end{cases}
\end{eqnarray}
The relic density of DM for $n < 1$ has the form
\begin{eqnarray}\label{approx relic NT3}
\Omega_{\chi,\, \text{NT}}^{(3)} \sim
\cfrac{m_{\chi} T_{\text{Reh}}^{\frac{(n+5)}{2}}}{2^{\frac{(1-n)}{2}}}\, \left(\cfrac{\left(m_{\phi}T_{\text{min}} \right)^{\frac{(1-n)}{2}} -\left(2m_{\chi}^{2}\right)^{\frac{(1-n)}{2}}}{m_{\chi}^{(1-n)}\,m_{\phi}^{\frac{(1-n)}{2}}} \right)\ .
\end{eqnarray}
The relic density of dark matter produced from the collision between two particles in thermal bath has the form
\begin{eqnarray}
\Omega_{\chi, \text{Th}} = && g_{\chi}^{2}\, g_{R}^{2}\, \cfrac{\lambda}{C^{\prime}}\, \cfrac{\zeta(3)^{2}}{\pi^{5}}\,  \left(\cfrac{30}{\pi^{2}g_{*}(T)}\right)^{(n+6)/4}\, \left(\cfrac{\pi^{2}}{30}\,g_{*}(T_{0})\right)^{3/4}  \left(\cfrac{\pi^{2}}{30}\,g_{*}(T_{\text{Reh}})\right)^{(n+1)/4} \nonumber \\
&& \times \left[\cfrac{m_{\chi}}{M^{n+2}}\,T_{\text{Reh}}^{n+1}\, \sqrt{3}M_{\text{pl}}\, \cfrac{T_{0}^{3}}{\rho_{c}} \right] \,\,\,\,\,\,\,\,\,\,\,\,\,\,\,\,\,\,\,\,\,\,\,\,\,\,\,\,\,\,\,\,\,\,\,\,\,\,\,\,\,\,\,\,\,\,\,\,\,\,\,\,\, \text{for}\; T_{\text{Reh}} > m_{\chi}. \nonumber \\
\end{eqnarray}
where $C^{\prime} = \frac{(2k_{1} - 3)(n+6) + 36}{8}$.

\section{Parameter space for cosmological models}\label{par_space}
In this section, we briefly discuss the microscopic models in which we can have  different kinds of temperature evolution depending upon the values of $k_{1}$ and $\omega$. Following ref.~\citep{Banerjee:2022fiw, Co:2020xaf}, for a rotating scalar field $\phi$ one can express the dissipation rate as $\Gamma_{\phi} \propto \bar{\phi}^{-2}$ for $m_{\phi} \gg T$. $\bar{\phi}$ redshifts as $a^{-3/2}$ for a $\phi^{2}$ potential gives $k_{1} = 3$, $\omega = 0$. For a scalar field $\phi$ oscillating near the minimum of the $\phi^{2}$ potential, the average dissipation rate  for $m_{\phi} \gg T$, $\Gamma_{\phi} \propto \bar{\phi}^{-1} \propto a^{3/2}$ gives $k_{1} = 3/2, \omega = 0$. If the scalar field oscillates around the minimum of the $\phi^{2}$ potential that gives $k_{1} = 0, \omega = 0$.

\subsubsection*{Case A : Increasing temperature during reheating}
\begin{figure}[h]
\centering
{\includegraphics[width=0.49\textwidth]{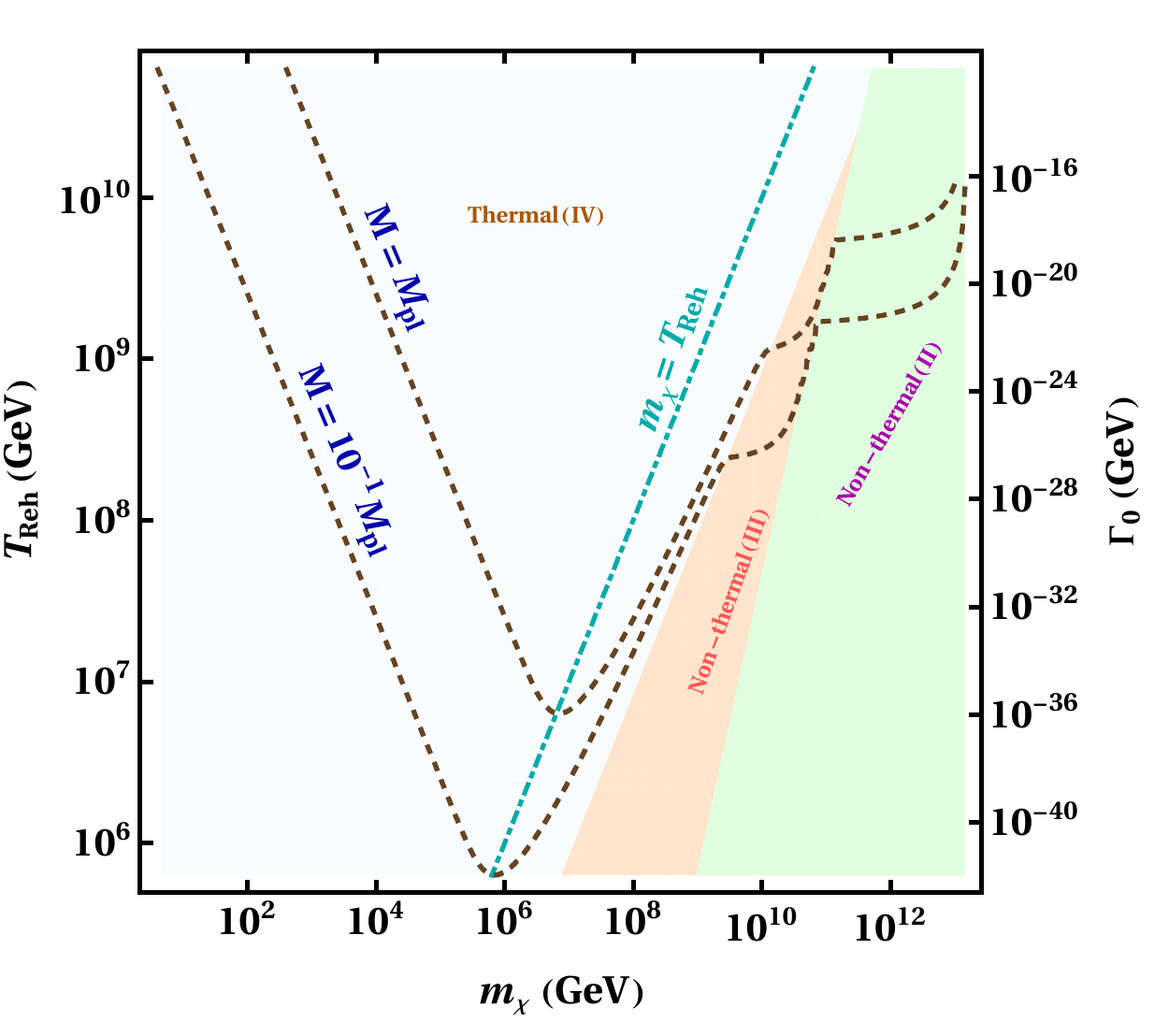}}
\hfill
{\includegraphics[width=0.49\textwidth]{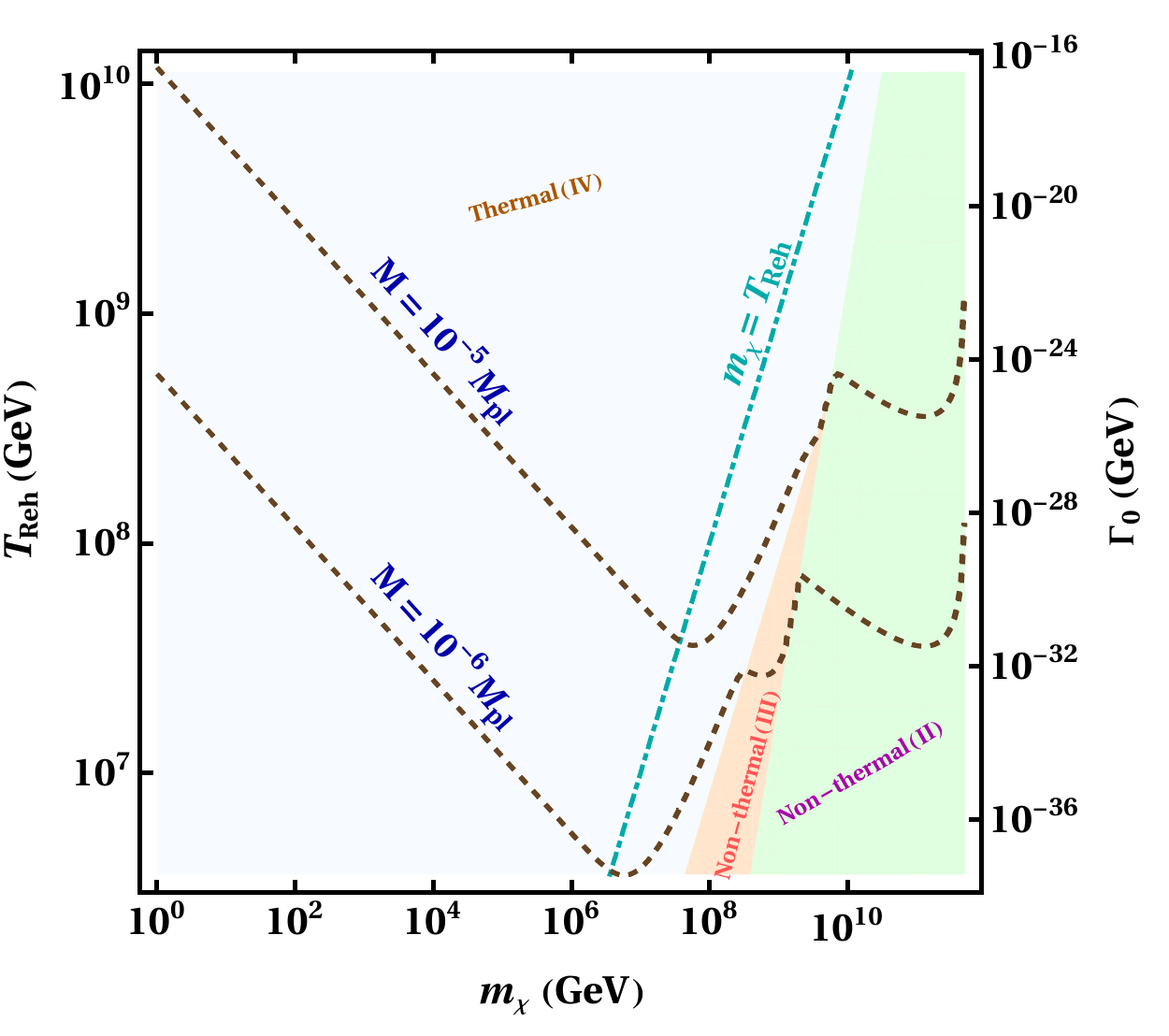}}
\caption{Different colored regions illustrate the parameter spaces in which various contributions to dark matter relic are dominant. Each dashed contour manifests the parameters that can elucidate the observed relic density of DM. We have displayed the specific value of $M$ associated with each contour. We set $k_{1} = 3.0, \omega = 0, H_{\text{end}} = 10^{12}$ GeV. We have fixed $n = 0$ and $m_{\phi} = 3 \times 10^{13}$ GeV in the left panel, and  $n$ = 2 and $m_{\phi} = 10^{12}$ GeV  in the right panel.}
\label{fig:contI}
\end{figure}

In Fig. (\ref{fig:contI}), we exhibit the contours of observed dark matter relic density i.e., $\Omega_{\chi}h^{2}$ = 0.12 inferred by PLANCK Collaboration \citep{Planck:2018vyg} in $T_{\text{Reh}} - m_{\chi}$ plane for n = 0\, (left) and n = 2\, (right).
In the regime $m_{\chi} < T_{\mathrm{Reh}}$, the blue-shaded region demonstrates the dominance of thermal contribution (thermal - IV) to the relic density of DM over other contributions.

The relic density of DM produced from the process thermal(IV) scales as $\Omega_{\chi} \sim m_{\chi} T_{\text{Reh}}^{n+1}$. As a result, to produce the observed DM, one needs to reduce $T_{\text{Reh}}$ with increasing DM mass. The constant relic density contour falls off with a slope of $1/(n+1)$ until the point where DM mass equals the reheating temperature i.e. $T_{\text{Reh}} = m_{\chi}$. \\
In the present scenario, $T_{\text{Reh}}$ being the maximum temperature attained by the thermal bath during the reheating, thermal production of DM is kinematically forbidden when $m_{\chi} > T_{\text{Reh}}$. Therefore, one should expect a sharp increase in reheating temperature along the line $m_{\chi} = T_{\text{Reh}}$. 
However, due to the tail of the modified Bessel function $K_{1}$ in Eq.~\eqref{thrate}, one gets a smooth increase in $T_{\text{Reh}}$ for $m_{\chi} > T_{\text{Reh}}$. 

In this regime where $m_{\chi} > T_{\text{Reh}}$, although the pair production of DM from the scattering of two thermal bath particles is kinematically disallowed, DM can still be produced from the collision of a bath particle with a high-energy particle. For $n = 2$, the relic density for the above non-thermal process scales as $\Omega_{\chi}h^{2} \sim m_{\chi} T_{\mathrm{Reh}}^{7/2}$. So as one moves towards higher dark matter mass one needs to decrease $T_{\mathrm{Reh}}$. 
In contrast for $n = 0$, the relic density scales as $\Omega_{\chi}h^{2} \sim \left(\left(m_{\phi}T_{\text{min}} \right)^{1/2} - \sqrt{2} m_{\chi}\right)\,T_{\text{Reh}}^{5/2}$ . In this case, the relic density of DM decreases with increasing DM mass for a fixed $T_{\text{Reh}}$. As a result, $T_{\text{Reh}}$ must increase to attain the observed relic density. These behaviors are visible in the orange-colored region in the figure.  
From Eq.~\eqref{nont3}, one can see, as $m_{\chi}^{2}$ approaches $\cfrac{m_{\phi}}{2}\, T_{\text{min}}$, DM number density drops very close to zero. Therefore, one needs a higher reheating temperature $T_{\text{Reh}}$ to obtain the observed relic density, illustrated by the sharp increase in $T_{\text{Reh}}$.

\begin{figure}[!h]
	\centering
	{\includegraphics[width=0.49\textwidth]{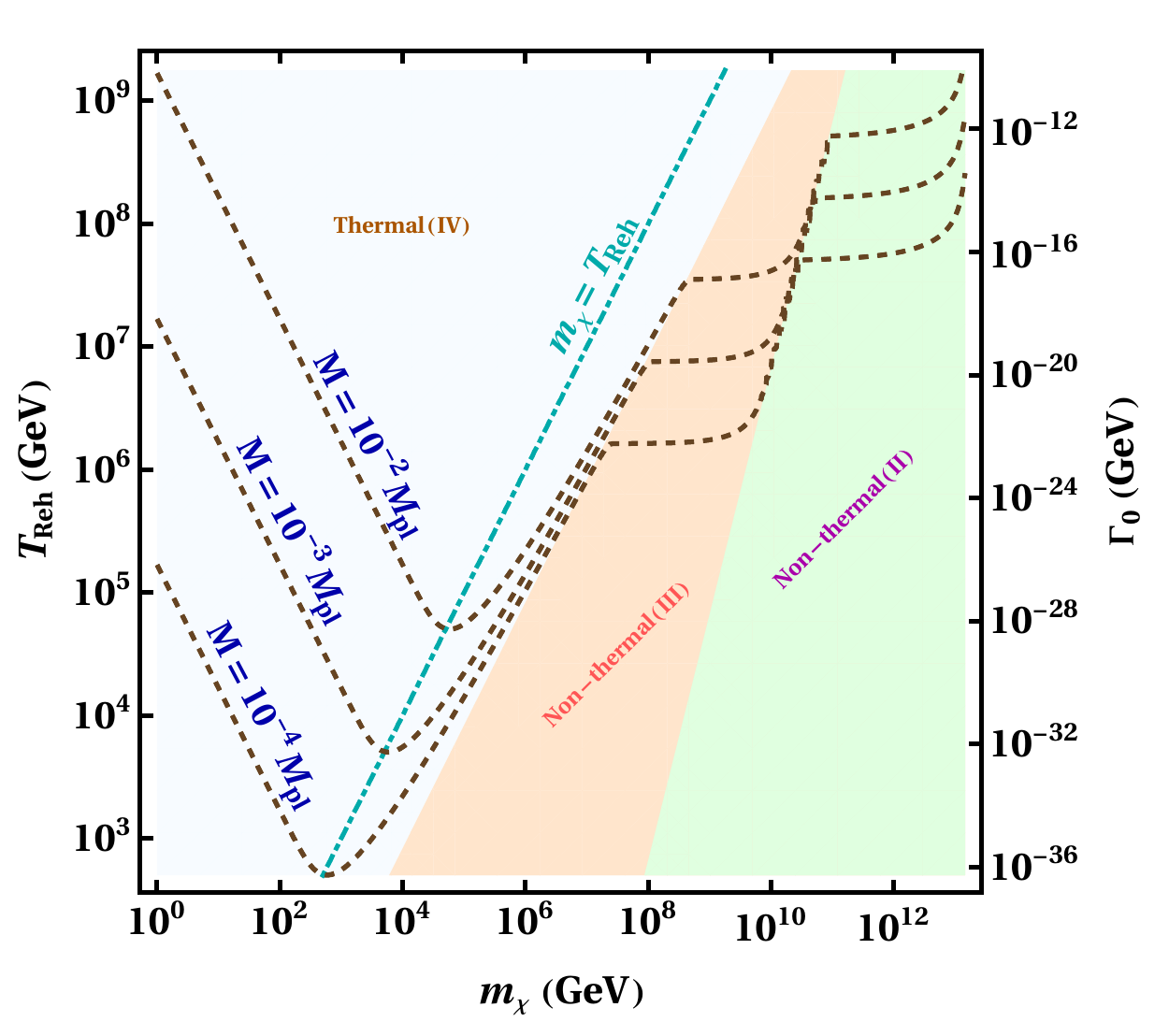}}
	\hfill
	{\includegraphics[width=0.49\textwidth]{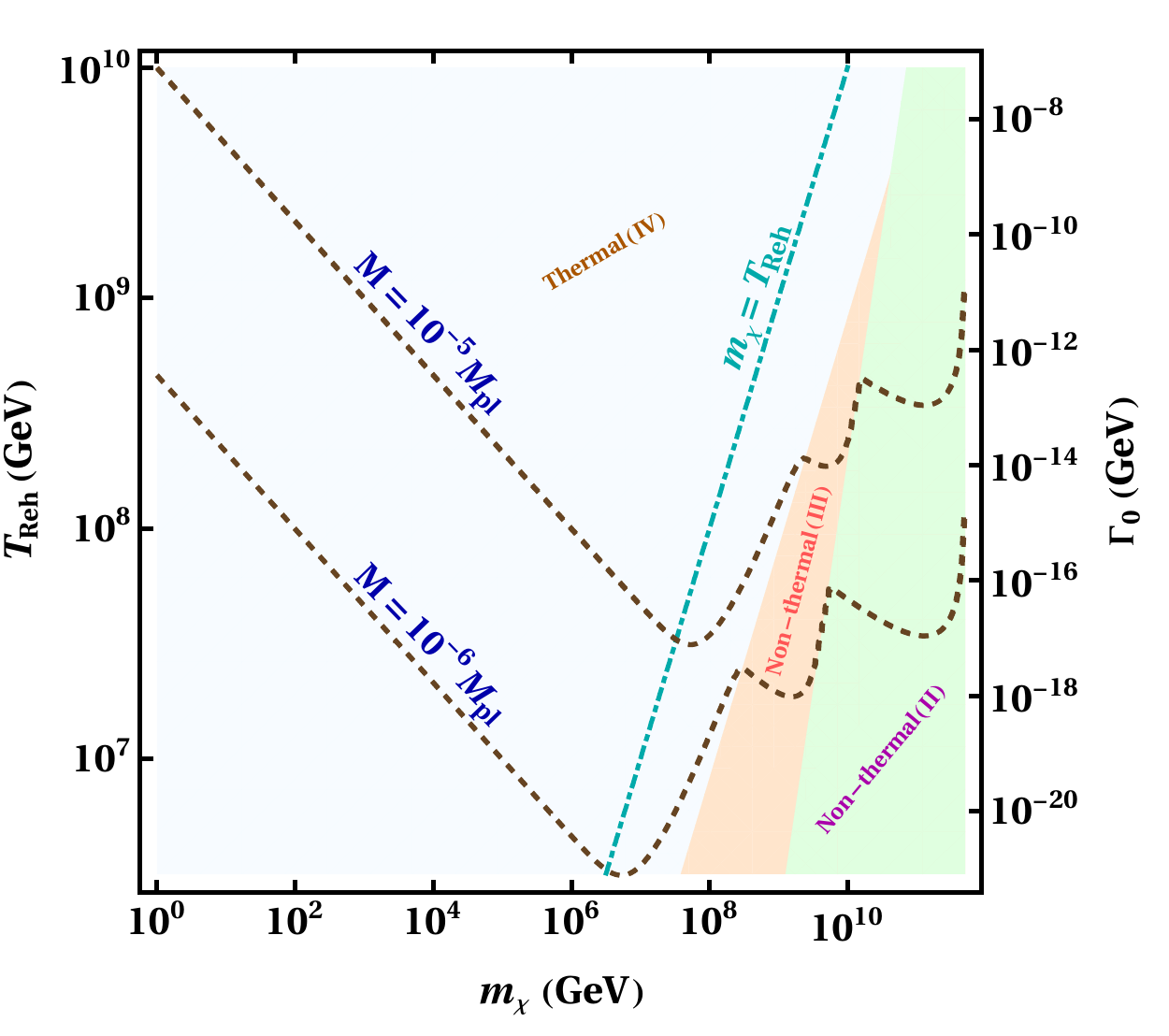}}
	\caption{
		Same as Fig. (\ref{fig:contI}) with $k_{1} = 3/2$.}
	\label{fig:contC}
\end{figure}
In the regime where $m_{\chi}^{2} > \frac{m_{\phi}}{2}T_{\text{min}}$, the exhaustive contribution in DM relic density comes from purely non-thermal processes, i.e., from the scattering between two high-energy particles either before or after thermalization. The green-shaded region in fig. (\ref{fig:contI}) delineates this non-thermal domination of the DM relic.
In this region, non-thermal (II) contributes significantly to the relic density compared to non-thermal (I). From Eq.~\eqref{Nt2approx}, one can see that the relic density of DM produced through non-thermal(II) scales as $\Omega_{\chi}h^{2} \sim m_{\chi} T_{\text{Reh}}^{4}$ for $n = 2$. As one goes to a larger DM mass $m_{\chi}$, one needs to reduce the reheating temperature. However for $n = 0$, the relic density of the dark matter has the expression $\Omega_{\chi}h^{2} \sim \left(m_{\phi}^{1/2} - (2m_{\chi})^{1/2} \right)^{2} T_{\text{Reh}}^{4}$. In a similar fashion as in the case of non-thermal (III), DM number density reduces with increasing dark matter mass, therefore, $T_{\text{Reh}}$ must increase for obtaining the observed relic density. From Eq.~\eqref{nont2}, it is visible that as the DM mass moves very close to $m_{\phi}/2$, due to the cancellation between $m_{\phi}/2$ and $m_{\chi}$, DM number density falls off drastically, demanding the $T_{\text{Reh}}$ to increase with increasing DM mass.

\subsubsection*{Case B: Constant temperature during reheating}

In Fig. (\ref{fig:contC}), we sketch the contours having observed dark matter relic density. In this case, the pattern of the constant relic density contours are almost similar to case A. But there are a couple of distinctions with case A. We discuss the differences in the next section.

\subsubsection*{Case C: Decreasing temperature during reheating}
\begin{figure}[h]
\centering
  {\includegraphics[width=0.49\textwidth]{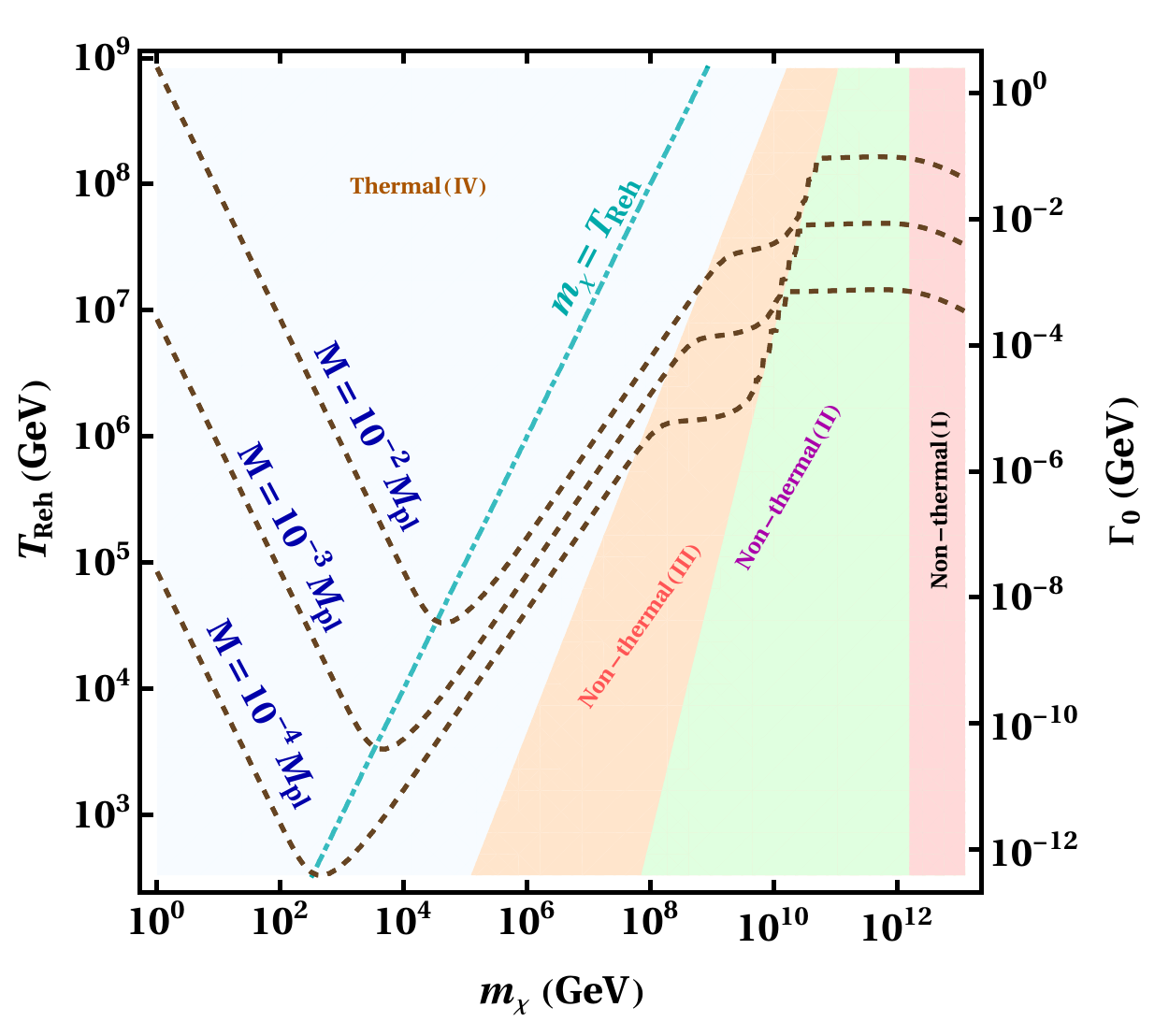}}
  \hfill
  {\includegraphics[width=0.49\textwidth]{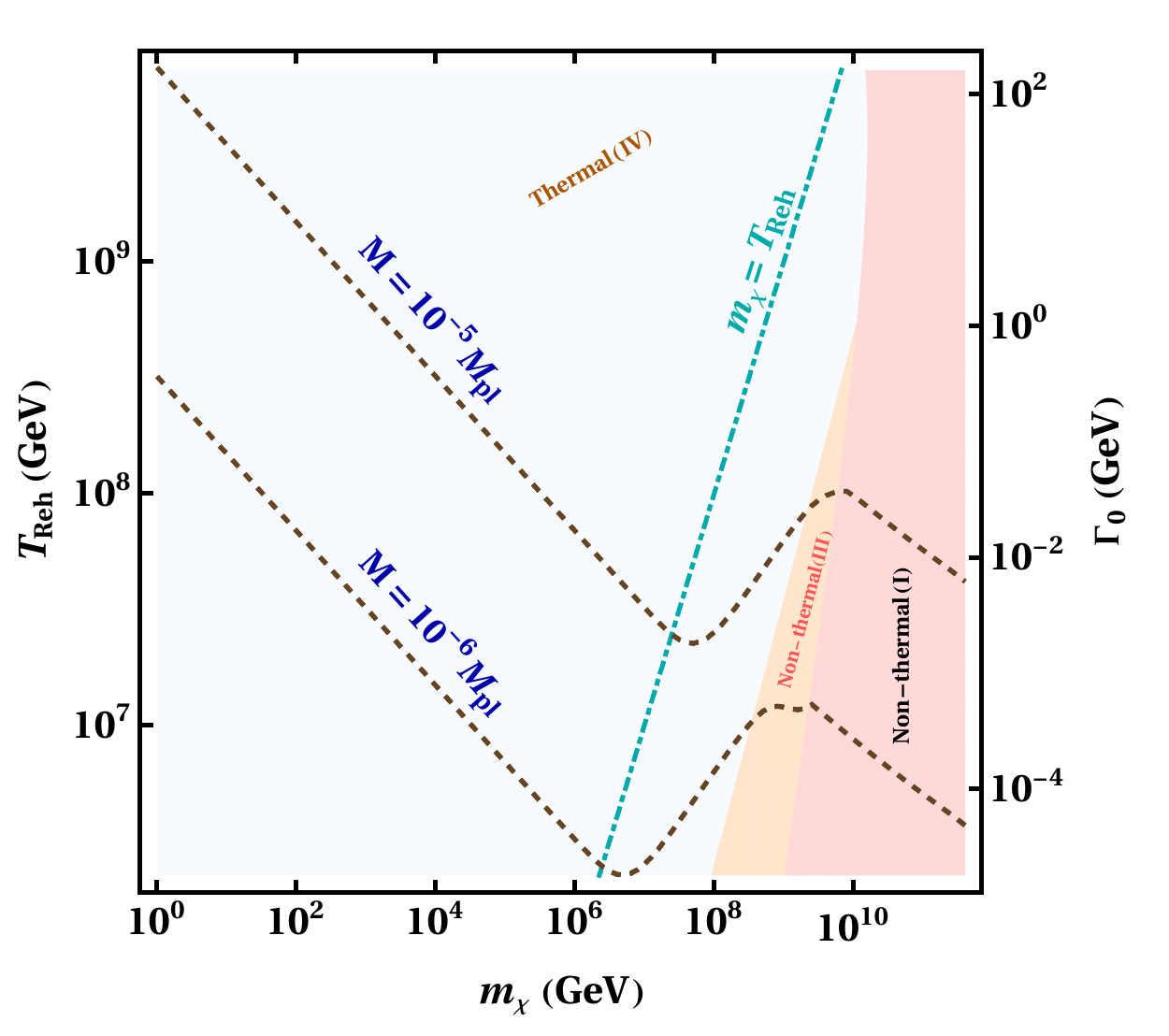}}
 \caption{
Same as Fig. (\ref{fig:contI}) with $k_{1} = 0$.}
  \label{fig:contD}
  \end{figure}
In Fig. (\ref{fig:contD}), we depict the contours of observed dark matter relic density. Although, the pattern of the contours looks similar to case A and case B. But there are a couple of differences with case A and case B. The differences are as follows,
$T_{\text{th}}$ is the minimum temperature during reheating for case A, whereas $T_{\text{Reh}}$ is the minimum temperature for case C. The temperature remains constant for case B, which implies  $T_{\text{Reh}}$($T_{\text{th}}$), can be considered as the minimum temperature of the bath during reheating.
Only for Case C, there exists a small region when DM mass is approximately close to $m_{\phi}/2$, i.e., $m_{\chi} \sim m_{\phi}/2$, the relic density is dominated by the DM produced from the scattering of high-energy particles before thermalization. As can be seen from the table (\ref{table:4}) that the high-energy decay products take longer time to thermalize for $k_{1} = 0$ as compared to non-zero $k_{1}$. So, these high-energy particles which are not yet thermalized get sufficient amount of time to produce the observe dark matter relic.
From the expression of $T_{\text{Reh}}$ given in Eq.~\eqref{Treh}, one can clearly see that $T_{\text{Reh}}$ depends on $z_{\text{Reh}}$, which in turn depends on $\Gamma_{0}$ and $k_{1}$ for fixed $H_{\text{end}}$.
For a fixed $\Gamma_{0}$ and DM mass to attain observed dark matter relic, one needs different $M$ for case B and case C compared to case A. 
\section{Comparison between various contributions in dark matter relic}\label{Comp_con}
 \begin{figure}[h]
\centering
  {\includegraphics[width=0.32\textwidth]{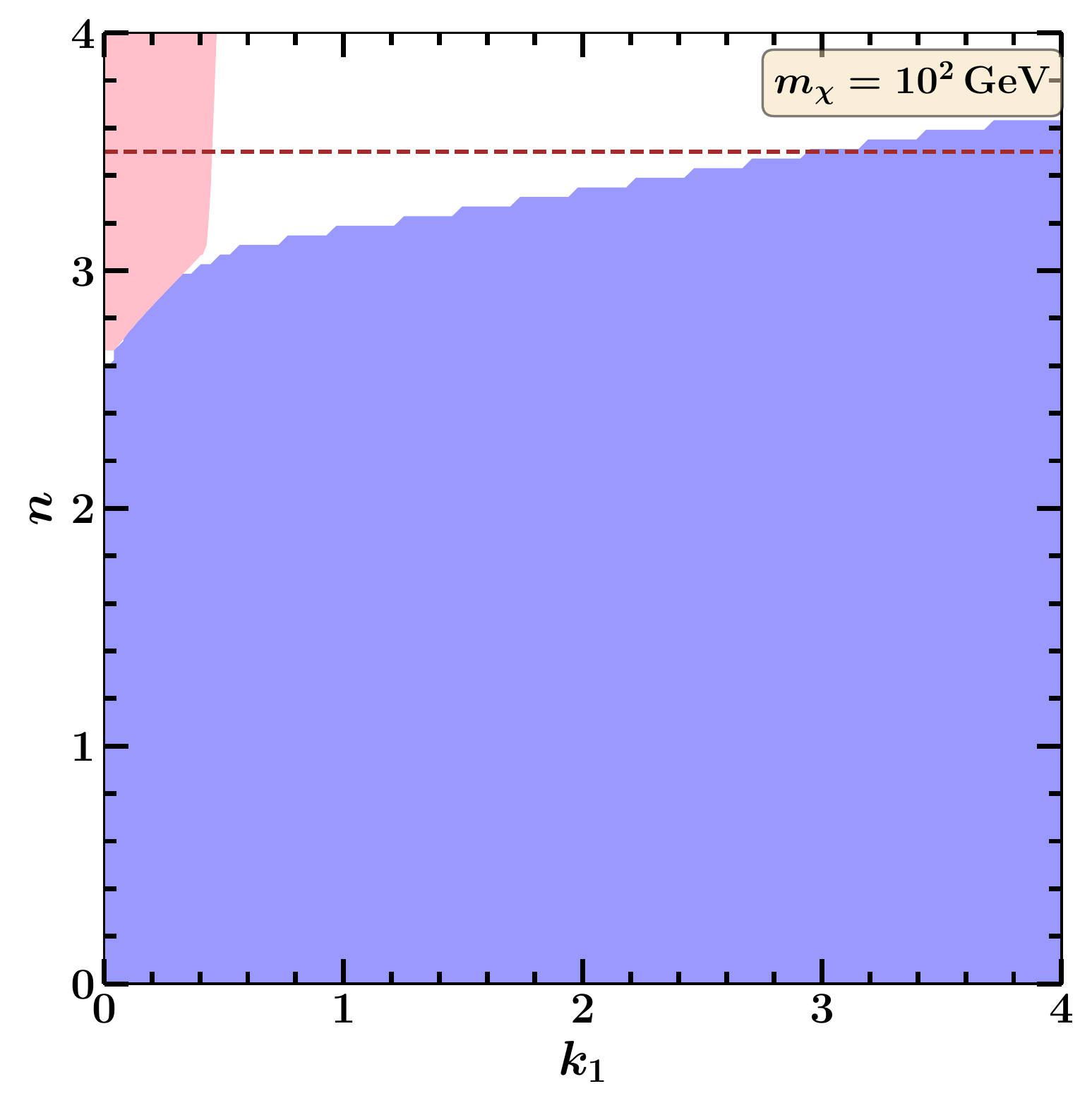}}
  \hfill
  {\includegraphics[width=0.32\textwidth]{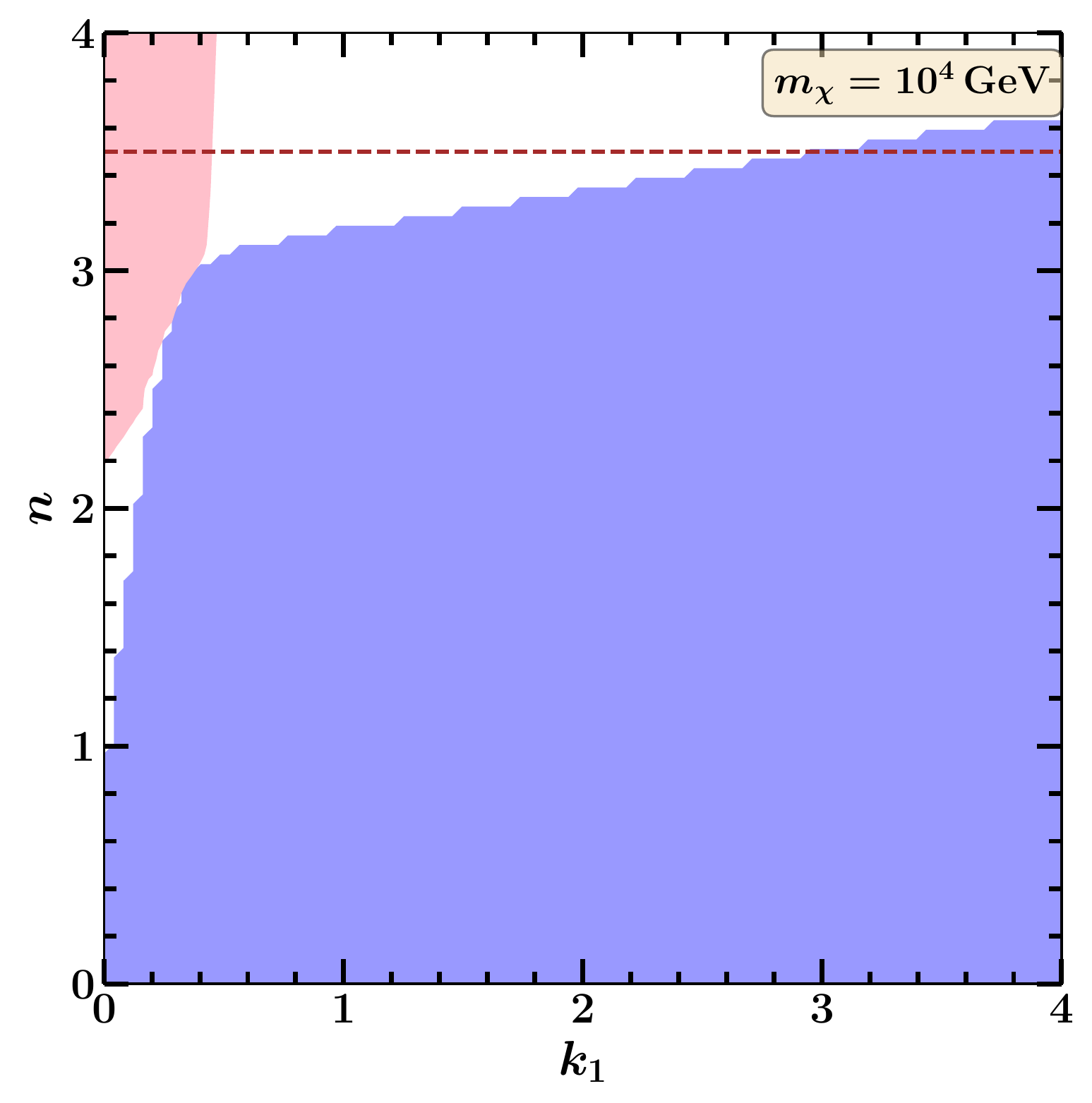}}
  \hfill
  {\includegraphics[width=0.32\textwidth]{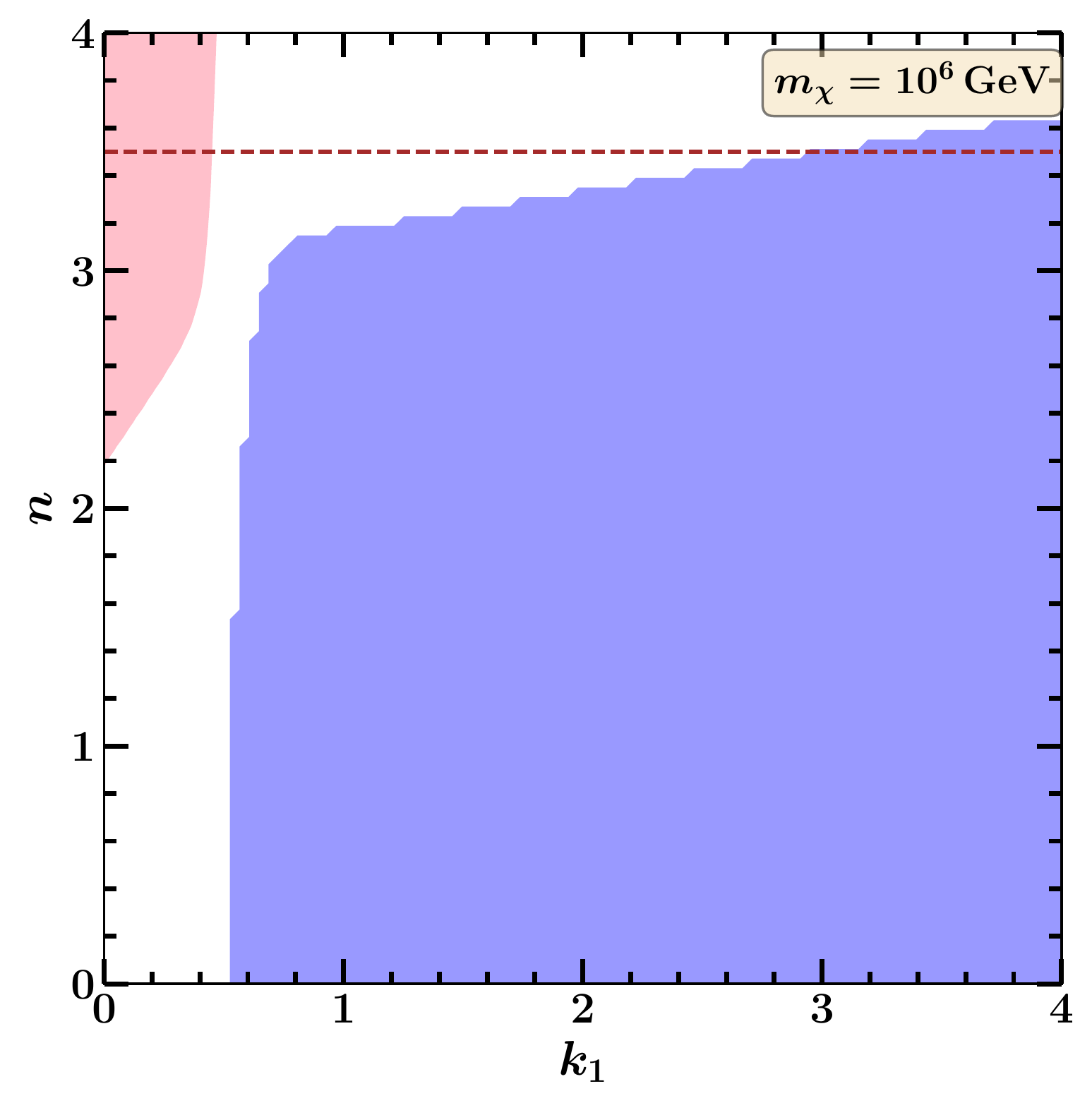}}
  \caption{We compare various contributions to DM relic density, and we have shown explicitly the regions where one contribution dominates over the other in $n - k_{1}$ parameter space. The left, middle and right panels are for different values of DM mass. In the plots we have set $\omega = 0, \Gamma_{0} = 10^{-15}\, \text{GeV}, M = 10^{-4} M_{\text{pl}}$. Blue colored region indicates that DM produced from the bath particles dominates the DM relic. In the non-thermal dominant region, pink-colored region illustrates that DM produced from the high-energy particles before thermalization dominates over the production after thermalization.}
  \label{fig:fc}
\end{figure} 
In Fig. (\ref{fig:fc}), we compare various contributions to DM relic density for fixed $m_{\chi}$, $\Gamma_{0}$, $H_{\text{end}}$, and $M$ while varying $n$ and $k_{1}$. The left, middle and right panels are for three different values of DM mass. First, we compare the relic density of DM produced from the thermal process and the non-thermal processes. We define the quantity
\begin{eqnarray}
f_{\text{th}} = \cfrac{\Omega_{\chi,\, \text{Th}}}{\Omega_{\chi,\, \text{NT}}^{(1)} + \Omega_{\chi,\, \text{NT}}^{(2)} + \Omega_{\chi,\, \text{NT}}^{(3)}}.
\end{eqnarray}
In the blue-colored region, $f_{\text{th}} > 1$ indicating the dominance of thermal relic of DM over non-thermal relic, i.e., DM relic produced from the scattering of two high-energy particles before and after thermalization, and the scattering of a high-energy particle with a thermal bath particle after thermalization, and in the rest of the parameter space, $f_{\text{th}} < 1$ connotes the non-thermal domination into DM relic. In this parameter space, there exists a subset of region where non-thermal contribution before thermalization dominates over non-thermal contributions after thermalization, measured by
\begin{eqnarray}
f_{\text{nt}} = \cfrac{\Omega_{\chi,\, \text{NT}}^{(1)}}{\Omega_{\chi,\, \text{NT}}^{(2)} + \Omega_{\chi,\, \text{NT}}^{(3)}}.
\end{eqnarray} 
In the pink-colored region $f_{\text{nt}} > 1$. From figure (\ref{fig:fc}), we observed that non-thermal contribution to DM relic is always dominant for $n \gtrsim 3.5$. For a higher DM mass, we even noticed non-thermal dominance when $k_{1} < 1$ for all values of $n$. This is because the thermal production of DM falls exponentially fast  when $m_{\chi}$ becomes larger than $T_{\text{max}}$ (maximum temperature during reheating). DM produced before the thermalization of the decay products contributes significantly when $n > 2$ with $k_{1} < 0.3$. These numerical results have also been verified with analytical results.

\section{Conclusions}\label{sec:conclusion}
After inflation, a meta-stable field that dominates the energy density of the universe dissipates its energy density via the production of high-energy particles. In general, this dissipation rate can vary with time. In this article, we have studied the impact of this time-dependent dissipation in the production of radiation after the end of inflation. The decay products of this meta-stable field can take a finite amount of time to reach thermal equilibrium with each other. We have studied the thermalization of these high-energy decay products of this meta-stable field in this non-standard cosmological scenario. We find that the decay products take shorter time to thermalize for non-zero $k_{1}$ compared to the case with $k_{1} = 0$. 

During reheating, the meta-stable field produces a shower of high-energy particles. These particles take some amount of time to dissipate their energy into low-energy particles (soft) to reach in equilibrium with the thermal bath. The temperature of the thermal bath varies differently as compared to the case of a constant dissipation. This unusual variation of temperature can affect the dark matter abundance during this epoch. 

In the standard picture, the particles in the thermal bath particles scatter with each other and produce dark matter. In addition to this, highly energetic particles can annihilate into pairs of dark matter. Scattering of these particles with thermal bath particles can also produce dark matter. We calculate the dark matter number density produced from these processes in various scenarios e.g., when the temperature remains constant, decreases, or increases during reheating. We have sketched the relic contours satisfying the observed dark matter abundance for $n$ = 0 and $n$ = 2, where $n$ is the power of the center of mass energy in dark matter production cross-section. The production of dark matter from these high-energy particles is important when the maximum temperature of the universe is smaller than the DM mass. We observed that DM production from the bath particles dominates when $m_{\chi} < T_{\text{max}}$. On the other side, when $m_{\chi} > T_{\text{max}}$ production of dark matter from the particles in the thermal bath gets kinematically suppressed. Dark matter still gets produced from the collision of a high-energy decay product and a particle in the thermal bath. This process gets kinematically blocked when $m_{\chi}^{2} > T_{\text{min}}m_{\phi}/2$. After the kinematical blocking of this process, the collision between two high-energy particles either before or after thermalization still can produce dark matter. We find that in the temperature decreasing scenario, dark matter produced from the high-energy particles before thermalization contributes significantly to dark matter relic density when $m_{\chi} \sim m_{\phi}/2$. Finally, we compare the relic density of DM produced from various production channels and find that DM production from non-thermal processes is always dominant for $n > 3$.

\section*{Acknowledgements}
The authors would like to thank Mustafa A. Amin, Avik Banerjee, and Marcos A. G. Garcia for carefully reading the manuscript and for their valuable comments. A.H. would like to thank MHRD, Government of India for the research fellowship. This research of D.C. is supported by an initiation grant IITK/PHY/2019413 at IIT Kanpur and by a DST-SERB grant SERB/CRG/2021/007579.

\appendix

\section{Phase Space distribution function of high-energy decay products}\label{Appendix:A}
The Boltzmann equation in terms of the phase-space distribution function of high-energy decay products
\begin{eqnarray}\label{BEQ}
\cfrac{df_{R}}{dt} - Hp \cfrac{df_{R}}{dp} & = & \mathcal{C}[f_{R}(p,\, t)] \nonumber \\
& = & \cfrac{4\pi^{2}}{p^{2}}\,(1 + \omega)\, n_{\phi}(t)\, \Gamma_{\phi}(t)\, \delta\left(p-\cfrac{m_{\phi}(t)}{2}\right).
\end{eqnarray}
The general solution of the Eq.~\eqref{BEQ} can be written as \citep{Ballesteros:2020adh}
\begin{eqnarray}
f_{R}(p, t) & = & \int_{t_{i}}^{t} dt^{\prime}\, \mathcal{C}\left[f_{R} \left( \cfrac{a(t)}{a(t^{\prime})}|p|,\, t^{\prime} \right)\right] \nonumber \\
& = & \int_{t_{\text{end}}}^{t} dt^{\prime}\, \cfrac{16\pi^{2}}{m_{\phi}^{2}(t^{\prime})}\,(1 + \omega)\,  n_{\phi}(t^{\prime})\, \Gamma_{\phi}(t^{\prime}) \delta\left(\frac{a(t)}{a(t^{\prime})}p - \cfrac{m_{\phi}(t^{\prime})}{2}\right) \nonumber \\ 
&=& \int_{t_{\text{end}}}^{t} dt^{\prime}\, \cfrac{32\pi^{2}}{m_{\phi}^{2}(t^{\prime})}\,\cfrac{(1 + \omega)}{|(3\omega - 1)|}\,  \cfrac{n_{\phi}(t^{\prime)}\, \Gamma_{\phi}(t^{\prime})}{m_{\phi}(\hat{t})\,H(\hat{t})}\, \delta(t^{\prime} - \hat{t})\,   \nonumber \\
& \simeq & \cfrac{8\pi^{2}}{m_{\phi}^{3}(\hat{t})}\,\cfrac{(2k_{1} + 3(1 + \omega) )}{|(3\omega - 1)|}\, n_{R}(t)\, \left(\cfrac{m_{\phi}(\hat{t})}{2p}\right)^{\frac{(3(1 - \omega)-2k_{1})}{2}}\, \theta\left(\cfrac{m_{\phi}(\hat{t})}{2}-p\right),
\end{eqnarray}
with $\frac{a(t)}{a(\hat{t})} = \left(\frac{m_{\phi}(t)}{2p}\right)^{\frac{1}{1 - 3\omega}}$ and $n_{R} = \frac{6(1+\omega)}{2k_{1} + 3(1+\omega)}n_{\phi}(t)\Gamma_{\phi}(t)t$ in the regime $\Gamma_{\phi}t \ll 1$.

\section{Reheating Temperature}
In this subsection, we derive the expression for the reheating temperature. Reheating concludes when the energy density of the meta-stable field is equal to the energy density of radiation i.e., $\rho_{\phi}(z_{\text{Reh}}) = \rho_{R}(z_{\text{Reh}})$.\\
The energy density of the meta-stable field has the form
\begin{eqnarray}
\rho_{\phi}(z) \approx \rho_{\phi, \text{end}}\, z^{-3(1 + \omega)} = 3\, M_{\text{pl}}^{2}\, H_{\text{end}}^{2}\, z^{-3(1 + \omega)}.
\end{eqnarray}
The energy density of radiation can be approximated as
\begin{eqnarray}
\rho_{R}(z) \approx \left(\cfrac{2(1 + \omega)}{2k_{1} - 3\omega + 5}\, \cfrac{\Gamma_{0}}{H_{\text{end}}}\, 3 M_{\text{pl}}^{2}\, H_{\text{end}}^{2}\right)\, z^{\frac{2k_{1} - 3(1 + \omega)}{2}}.
\end{eqnarray}
Reheating condition i.e. $\rho_{\phi}(z_{\text{Reh}}) = \rho_{R}(z_{\text{Reh}})$ gives
\begin{eqnarray}
z_{\text{Reh}} = \cfrac{\left(3 M_{\text{pl}}^{2} H_{\text{end}}^{2}\right)^{\frac{2}{2k_{1} + 3(1 + \omega)}}}{\left(\cfrac{2(1 + \omega)}{2k_{1} - 3\omega + 5}\, \cfrac{\Gamma_{0}}{H_{\text{end}}}\, 3 M_{\text{pl}}^{2}\, H_{\text{end}}^{2} \right)^{\frac{2}{2k_{1} + 3(1 + \omega)}}}.
\end{eqnarray}

The temperature evolution of the universe after thermalization until reheating can be written as a function of redshift $z$ and is given by,
\begin{eqnarray}
T(z) & = & \left( \cfrac{30}{\pi^{2}\, g_{*}(T)} \right)^{1/4}\, \rho_{R}(z)^{1/4} \nonumber\\
& = & \left( \cfrac{30}{\pi^{2}\, g_{*}(T)} \right)^{1/4}\, \left(\cfrac{2(1 + \omega)}{2k_{1} - 3\omega + 5}\, \cfrac{\Gamma_{0}}{H_{\text{end}}}\, 3 M_{\text{pl}}^{2}\, H_{\text{end}}^{2}\right)^{1/4}\, z^{\frac{2k_{1} - 3(1 + \omega)}{8}}.
\end{eqnarray}

Using the expression of $z_{\text{Reh}}$ we can write the reheating temperature
\begin{eqnarray}
T_{\text{Reh}} &=& \left( \cfrac{30}{\pi^{2}\, g_{*}(T_{\text{Reh}})} \right)^{1/4}\, \left(\cfrac{2(1 + \omega)}{2k_{1} - 3\omega + 5}\, \cfrac{\Gamma_{0}}{H_{\text{end}}}\, 3 M_{\text{pl}}^{2}\, H_{\text{end}}^{2}\right)^{1/4}\, z_{\text{Reh}}^{\frac{2k_{1} - 3(1 + \omega)}{8}} \nonumber \\
& = & \left( \cfrac{30}{\pi^{2}\, g_{*}(T_{\text{Reh}})} \right)^{1/4}\, \cfrac{(3 M_{\text{pl}}^{2}\, H_{\text{end}}^{2})^{1/4}}{z_{\text{Reh}}^{3(1+\omega)/4}}.
\end{eqnarray}

\newpage

\bibliographystyle{JHEP}
\bibliography{NTDM}

\providecommand{\href}[2]{#2}\begingroup\raggedright\begin{thebibliography}{10}

\bibitem{Allahverdi:2002nb}
R.~Allahverdi and M.~Drees, \emph{{Production of massive stable particles in
  inflaton decay}},
  \href{https://doi.org/10.1103/PhysRevLett.89.091302}{\emph{Phys. Rev. Lett.}
  {\bfseries 89} (2002) 091302}
  [\href{https://arxiv.org/abs/hep-ph/0203118}{{\ttfamily hep-ph/0203118}}].

\bibitem{Allahverdi:2010xz}
R.~Allahverdi, R.~Brandenberger, F.-Y.~Cyr-Racine and A.~Mazumdar,
  \emph{{Reheating in Inflationary Cosmology: Theory and Applications}},
  \href{https://doi.org/10.1146/annurev.nucl.012809.104511}{\emph{Ann. Rev.
  Nucl. Part. Sci.} {\bfseries 60} (2010) 27}
  [\href{https://arxiv.org/abs/1001.2600}{{\ttfamily 1001.2600}}].

\bibitem{Moroi:2002rd}
T.~Moroi and T.~Takahashi, \emph{{Cosmic density perturbations from late
  decaying scalar condensations}},
  \href{https://doi.org/10.1103/PhysRevD.66.063501}{\emph{Phys. Rev. D}
  {\bfseries 66} (2002) 063501}
  [\href{https://arxiv.org/abs/hep-ph/0206026}{{\ttfamily hep-ph/0206026}}].

\bibitem{Lahanas:2011tk}
A.B.~Lahanas, \emph{{Dilaton dominance in the early Universe dilutes Dark
  Matter relic abundances}},
  \href{https://doi.org/10.1103/PhysRevD.83.103523}{\emph{Phys. Rev. D}
  {\bfseries 83} (2011) 103523}
  [\href{https://arxiv.org/abs/1102.4277}{{\ttfamily 1102.4277}}].

\bibitem{Moroi:1999zb}
T.~Moroi and L.~Randall, \emph{{Wino cold dark matter from anomaly mediated
  SUSY breaking}},
  \href{https://doi.org/10.1016/S0550-3213(99)00748-8}{\emph{Nucl. Phys. B}
  {\bfseries 570} (2000) 455}
  [\href{https://arxiv.org/abs/hep-ph/9906527}{{\ttfamily hep-ph/9906527}}].

\bibitem{Starobinsky:1994bd}
A.A.~Starobinsky and J.~Yokoyama, \emph{{Equilibrium state of a selfinteracting
  scalar field in the De Sitter background}},
  \href{https://doi.org/10.1103/PhysRevD.50.6357}{\emph{Phys. Rev. D}
  {\bfseries 50} (1994) 6357}
  [\href{https://arxiv.org/abs/astro-ph/9407016}{{\ttfamily
  astro-ph/9407016}}].

\bibitem{Dine:1995uk}
M.~Dine, L.~Randall and S.D.~Thomas, \emph{{Supersymmetry breaking in the early
  universe}}, \href{https://doi.org/10.1103/PhysRevLett.75.398}{\emph{Phys.
  Rev. Lett.} {\bfseries 75} (1995) 398}
  [\href{https://arxiv.org/abs/hep-ph/9503303}{{\ttfamily hep-ph/9503303}}].

\bibitem{Acharya:2008bk}
B.S.~Acharya, P.~Kumar, K.~Bobkov, G.~Kane, J.~Shao and S.~Watson,
  \emph{{Non-thermal Dark Matter and the Moduli Problem in String Frameworks}},
  \href{https://doi.org/10.1088/1126-6708/2008/06/064}{\emph{JHEP} {\bfseries
  06} (2008) 064} [\href{https://arxiv.org/abs/0804.0863}{{\ttfamily
  0804.0863}}].

\bibitem{Kane:2015jia}
G.~Kane, K.~Sinha and S.~Watson, \emph{{Cosmological Moduli and the
  Post-Inflationary Universe: A Critical Review}},
  \href{https://doi.org/10.1142/S0218271815300220}{\emph{Int. J. Mod. Phys. D}
  {\bfseries 24} (2015) 1530022}
  [\href{https://arxiv.org/abs/1502.07746}{{\ttfamily 1502.07746}}].

\bibitem{Chung:1998rq}
D.J.H.~Chung, E.W.~Kolb and A.~Riotto, \emph{{Production of massive particles
  during reheating}},
  \href{https://doi.org/10.1103/PhysRevD.60.063504}{\emph{Phys. Rev. D}
  {\bfseries 60} (1999) 063504}
  [\href{https://arxiv.org/abs/hep-ph/9809453}{{\ttfamily hep-ph/9809453}}].

\bibitem{Giudice:2000ex}
G.F.~Giudice, E.W.~Kolb and A.~Riotto, \emph{{Largest temperature of the
  radiation era and its cosmological implications}},
  \href{https://doi.org/10.1103/PhysRevD.64.023508}{\emph{Phys. Rev. D}
  {\bfseries 64} (2001) 023508}
  [\href{https://arxiv.org/abs/hep-ph/0005123}{{\ttfamily hep-ph/0005123}}].

\bibitem{Allahverdi:2002pu}
R.~Allahverdi and M.~Drees, \emph{{Thermalization after inflation and
  production of massive stable particles}},
  \href{https://doi.org/10.1103/PhysRevD.66.063513}{\emph{Phys. Rev. D}
  {\bfseries 66} (2002) 063513}
  [\href{https://arxiv.org/abs/hep-ph/0205246}{{\ttfamily hep-ph/0205246}}].

\bibitem{Allahverdi:2002ap}
R.~Allahverdi and M.~Drees, \emph{{Heavy particle production during
  reheating}},  in \emph{{10th International Conference on Supersymmetry and
  Unification of Fundamental Interactions (SUSY02)}}, pp.~1183--1189, 10, 2002
  [\href{https://arxiv.org/abs/hep-ph/0210432}{{\ttfamily hep-ph/0210432}}].

\bibitem{Kane:2015qea}
G.L.~Kane, P.~Kumar, B.D.~Nelson and B.~Zheng, \emph{{Dark matter production
  mechanisms with a nonthermal cosmological history: A classification}},
  \href{https://doi.org/10.1103/PhysRevD.93.063527}{\emph{Phys. Rev. D}
  {\bfseries 93} (2016) 063527}
  [\href{https://arxiv.org/abs/1502.05406}{{\ttfamily 1502.05406}}].

\bibitem{Garcia:2017tuj}
M.A.G.~Garcia, Y.~Mambrini, K.A.~Olive and M.~Peloso, \emph{{Enhancement of the
  Dark Matter Abundance Before Reheating: Applications to Gravitino Dark
  Matter}}, \href{https://doi.org/10.1103/PhysRevD.96.103510}{\emph{Phys. Rev.
  D} {\bfseries 96} (2017) 103510}
  [\href{https://arxiv.org/abs/1709.01549}{{\ttfamily 1709.01549}}].

\bibitem{Drees:2017iod}
M.~Drees and F.~Hajkarim, \emph{{Dark Matter Production in an Early Matter
  Dominated Era}},
  \href{https://doi.org/10.1088/1475-7516/2018/02/057}{\emph{JCAP} {\bfseries
  02} (2018) 057} [\href{https://arxiv.org/abs/1711.05007}{{\ttfamily
  1711.05007}}].

\bibitem{Drees:2018dsj}
M.~Drees and F.~Hajkarim, \emph{{Neutralino Dark Matter in Scenarios with Early
  Matter Domination}},
  \href{https://doi.org/10.1007/JHEP12(2018)042}{\emph{JHEP} {\bfseries 12}
  (2018) 042} [\href{https://arxiv.org/abs/1808.05706}{{\ttfamily
  1808.05706}}].

\bibitem{Allahverdi:2018aux}
R.~Allahverdi and J.K.~Osi\'nski, \emph{{Nonthermal dark matter from modified
  early matter domination}},
  \href{https://doi.org/10.1103/PhysRevD.99.083517}{\emph{Phys. Rev. D}
  {\bfseries 99} (2019) 083517}
  [\href{https://arxiv.org/abs/1812.10522}{{\ttfamily 1812.10522}}].

\bibitem{Arias:2022qjt}
P.~Arias, N.~Bernal, J.K.~Osi\'nski and L.~Roszkowski, \emph{{Dark matter
  axions in the early universe with a period of increasing temperature}},
  \href{https://doi.org/10.1088/1475-7516/2023/05/028}{\emph{JCAP} {\bfseries
  05} (2023) 028} [\href{https://arxiv.org/abs/2207.07677}{{\ttfamily
  2207.07677}}].

\bibitem{Bernal:2022wck}
N.~Bernal and Y.~Xu, \emph{{WIMPs during reheating}},
  \href{https://doi.org/10.1088/1475-7516/2022/12/017}{\emph{JCAP} {\bfseries
  12} (2022) 017} [\href{https://arxiv.org/abs/2209.07546}{{\ttfamily
  2209.07546}}].

\bibitem{Bhattiprolu:2022sdd}
P.N.~Bhattiprolu, G.~Elor, R.~McGehee and A.~Pierce, \emph{{Freezing-in
  hadrophilic dark matter at low reheating temperatures}},
  \href{https://doi.org/10.1007/JHEP01(2023)128}{\emph{JHEP} {\bfseries 01}
  (2023) 128} [\href{https://arxiv.org/abs/2210.15653}{{\ttfamily
  2210.15653}}].

\bibitem{Chowdhury:2018tzw}
D.~Chowdhury, E.~Dudas, M.~Dutra and Y.~Mambrini, \emph{{Moduli Portal Dark
  Matter}}, \href{https://doi.org/10.1103/PhysRevD.99.095028}{\emph{Phys. Rev.
  D} {\bfseries 99} (2019) 095028}
  [\href{https://arxiv.org/abs/1811.01947}{{\ttfamily 1811.01947}}].

\bibitem{Banerjee:2019asa}
A.~Banerjee, G.~Bhattacharyya, D.~Chowdhury and Y.~Mambrini, \emph{{Dark matter
  seeping through dynamic gauge kinetic mixing}},
  \href{https://doi.org/10.1088/1475-7516/2019/12/009}{\emph{JCAP} {\bfseries
  12} (2019) 009} [\href{https://arxiv.org/abs/1905.11407}{{\ttfamily
  1905.11407}}].

\bibitem{Garcia:2020eof}
M.A.G.~Garcia, K.~Kaneta, Y.~Mambrini and K.A.~Olive, \emph{{Reheating and
  Post-inflationary Production of Dark Matter}},
  \href{https://doi.org/10.1103/PhysRevD.101.123507}{\emph{Phys. Rev. D}
  {\bfseries 101} (2020) 123507}
  [\href{https://arxiv.org/abs/2004.08404}{{\ttfamily 2004.08404}}].

\bibitem{Garcia:2020wiy}
M.A.G.~Garcia, K.~Kaneta, Y.~Mambrini and K.A.~Olive, \emph{{Inflaton
  Oscillations and Post-Inflationary Reheating}},
  \href{https://doi.org/10.1088/1475-7516/2021/04/012}{\emph{JCAP} {\bfseries
  04} (2021) 012} [\href{https://arxiv.org/abs/2012.10756}{{\ttfamily
  2012.10756}}].

\bibitem{Garcia:2018wtq}
M.A.G.~Garcia and M.A.~Amin, \emph{{Prethermalization production of dark
  matter}}, \href{https://doi.org/10.1103/PhysRevD.98.103504}{\emph{Phys. Rev.
  D} {\bfseries 98} (2018) 103504}
  [\href{https://arxiv.org/abs/1806.01865}{{\ttfamily 1806.01865}}].

\bibitem{Harigaya:2014waa}
K.~Harigaya, M.~Kawasaki, K.~Mukaida and M.~Yamada, \emph{{Dark Matter
  Production in Late Time Reheating}},
  \href{https://doi.org/10.1103/PhysRevD.89.083532}{\emph{Phys. Rev. D}
  {\bfseries 89} (2014) 083532}
  [\href{https://arxiv.org/abs/1402.2846}{{\ttfamily 1402.2846}}].

\bibitem{Harigaya:2019tzu}
K.~Harigaya, K.~Mukaida and M.~Yamada, \emph{{Dark Matter Production during the
  Thermalization Era}},
  \href{https://doi.org/10.1007/JHEP07(2019)059}{\emph{JHEP} {\bfseries 07}
  (2019) 059} [\href{https://arxiv.org/abs/1901.11027}{{\ttfamily
  1901.11027}}].

\bibitem{Drees:2021lbm}
M.~Drees and B.~Najjari, \emph{{Energy spectrum of thermalizing high energy
  decay products in the early universe}},
  \href{https://doi.org/10.1088/1475-7516/2021/10/009}{\emph{JCAP} {\bfseries
  10} (2021) 009} [\href{https://arxiv.org/abs/2105.01935}{{\ttfamily
  2105.01935}}].

\bibitem{Drees:2022vvn}
M.~Drees and B.~Najjari, \emph{{Multi-Species Thermalization Cascade of
  Energetic Particles in the Early Universe}},
  \href{https://arxiv.org/abs/2205.07741}{{\ttfamily 2205.07741}}.

\bibitem{Mukaida:2022bbo}
K.~Mukaida and M.~Yamada, \emph{{Cascades of high-energy SM particles in the
  primordial thermal plasma}},
  \href{https://doi.org/10.1007/JHEP10(2022)116}{\emph{JHEP} {\bfseries 10}
  (2022) 116} [\href{https://arxiv.org/abs/2208.11708}{{\ttfamily
  2208.11708}}].

\bibitem{Davidson:2000er}
S.~Davidson and S.~Sarkar, \emph{{Thermalization after inflation}},
  \href{https://doi.org/10.1088/1126-6708/2000/11/012}{\emph{JHEP} {\bfseries
  11} (2000) 012} [\href{https://arxiv.org/abs/hep-ph/0009078}{{\ttfamily
  hep-ph/0009078}}].

\bibitem{Mukaida:2015ria}
K.~Mukaida and M.~Yamada, \emph{{Thermalization Process after Inflation and
  Effective Potential of Scalar Field}},
  \href{https://doi.org/10.1088/1475-7516/2016/02/003}{\emph{JCAP} {\bfseries
  02} (2016) 003} [\href{https://arxiv.org/abs/1506.07661}{{\ttfamily
  1506.07661}}].

\bibitem{Kurkela:2011ti}
A.~Kurkela and G.D.~Moore, \emph{{Thermalization in Weakly Coupled Nonabelian
  Plasmas}}, \href{https://doi.org/10.1007/JHEP12(2011)044}{\emph{JHEP}
  {\bfseries 12} (2011) 044} [\href{https://arxiv.org/abs/1107.5050}{{\ttfamily
  1107.5050}}].

\bibitem{Harigaya:2013vwa}
K.~Harigaya and K.~Mukaida, \emph{{Thermalization after/during Reheating}},
  \href{https://doi.org/10.1007/JHEP05(2014)006}{\emph{JHEP} {\bfseries 05}
  (2014) 006} [\href{https://arxiv.org/abs/1312.3097}{{\ttfamily 1312.3097}}].

\bibitem{Mukaida:2012qn}
K.~Mukaida and K.~Nakayama, \emph{{Dynamics of oscillating scalar field in
  thermal environment}},
  \href{https://doi.org/10.1088/1475-7516/2013/01/017}{\emph{JCAP} {\bfseries
  01} (2013) 017} [\href{https://arxiv.org/abs/1208.3399}{{\ttfamily
  1208.3399}}].

\bibitem{Co:2020xaf}
R.T.~Co, E.~Gonzalez and K.~Harigaya, \emph{{Increasing Temperature toward the
  Completion of Reheating}},
  \href{https://doi.org/10.1088/1475-7516/2020/11/038}{\emph{JCAP} {\bfseries
  11} (2020) 038} [\href{https://arxiv.org/abs/2007.04328}{{\ttfamily
  2007.04328}}].

\bibitem{Ai:2021gtg}
W.-Y.~Ai, M.~Drewes, D.~Glavan and J.~Hajer, \emph{{Oscillating scalar
  dissipating in a medium}},
  \href{https://doi.org/10.1007/JHEP11(2021)160}{\emph{JHEP} {\bfseries 11}
  (2021) 160} [\href{https://arxiv.org/abs/2108.00254}{{\ttfamily
  2108.00254}}].

\bibitem{Wang:2022mvv}
Z.-L.~Wang and W.-Y.~Ai, \emph{{Dissipation of oscillating scalar backgrounds
  in an FLRW universe}},
  \href{https://doi.org/10.1007/JHEP11(2022)075}{\emph{JHEP} {\bfseries 11}
  (2022) 075} [\href{https://arxiv.org/abs/2202.08218}{{\ttfamily
  2202.08218}}].

\bibitem{Hall:2009bx}
L.J.~Hall, K.~Jedamzik, J.~March-Russell and S.M.~West, \emph{{Freeze-In
  Production of FIMP Dark Matter}},
  \href{https://doi.org/10.1007/JHEP03(2010)080}{\emph{JHEP} {\bfseries 03}
  (2010) 080} [\href{https://arxiv.org/abs/0911.1120}{{\ttfamily 0911.1120}}].

\bibitem{Hall:2010jx}
L.J.~Hall, J.~March-Russell and S.M.~West, \emph{{A Unified Theory of Matter
  Genesis: Asymmetric Freeze-In}},
  \href{https://arxiv.org/abs/1010.0245}{{\ttfamily 1010.0245}}.

\bibitem{Cheung:2011nn}
C.~Cheung, G.~Elor and L.~Hall, \emph{{Gravitino Freeze-In}},
  \href{https://doi.org/10.1103/PhysRevD.84.115021}{\emph{Phys. Rev. D}
  {\bfseries 84} (2011) 115021}
  [\href{https://arxiv.org/abs/1103.4394}{{\ttfamily 1103.4394}}].

\bibitem{Mambrini:2013iaa}
Y.~Mambrini, K.A.~Olive, J.~Quevillon and B.~Zaldivar, \emph{{Gauge Coupling
  Unification and Nonequilibrium Thermal Dark Matter}},
  \href{https://doi.org/10.1103/PhysRevLett.110.241306}{\emph{Phys. Rev. Lett.}
  {\bfseries 110} (2013) 241306}
  [\href{https://arxiv.org/abs/1302.4438}{{\ttfamily 1302.4438}}].

\bibitem{Mambrini:2015vna}
Y.~Mambrini, N.~Nagata, K.A.~Olive, J.~Quevillon and J.~Zheng, \emph{{Dark
  matter and gauge coupling unification in nonsupersymmetric SO(10) grand
  unified models}},
  \href{https://doi.org/10.1103/PhysRevD.91.095010}{\emph{Phys. Rev. D}
  {\bfseries 91} (2015) 095010}
  [\href{https://arxiv.org/abs/1502.06929}{{\ttfamily 1502.06929}}].

\bibitem{Mambrini:2016dca}
Y.~Mambrini, N.~Nagata, K.A.~Olive and J.~Zheng, \emph{{Vacuum Stability and
  Radiative Electroweak Symmetry Breaking in an SO(10) Dark Matter Model}},
  \href{https://doi.org/10.1103/PhysRevD.93.111703}{\emph{Phys. Rev. D}
  {\bfseries 93} (2016) 111703}
  [\href{https://arxiv.org/abs/1602.05583}{{\ttfamily 1602.05583}}].

\bibitem{Dudas:2017kfz}
E.~Dudas, T.~Gherghetta, Y.~Mambrini and K.A.~Olive, \emph{{Inflation and
  High-Scale Supersymmetry with an EeV Gravitino}},
  \href{https://doi.org/10.1103/PhysRevD.96.115032}{\emph{Phys. Rev. D}
  {\bfseries 96} (2017) 115032}
  [\href{https://arxiv.org/abs/1710.07341}{{\ttfamily 1710.07341}}].

\bibitem{Garny:2017kha}
M.~Garny, A.~Palessandro, M.~Sandora and M.S.~Sloth, \emph{{Theory and
  Phenomenology of Planckian Interacting Massive Particles as Dark Matter}},
  \href{https://doi.org/10.1088/1475-7516/2018/02/027}{\emph{JCAP} {\bfseries
  02} (2018) 027} [\href{https://arxiv.org/abs/1709.09688}{{\ttfamily
  1709.09688}}].

\bibitem{Benakli:2017whb}
K.~Benakli, Y.~Chen, E.~Dudas and Y.~Mambrini, \emph{{Minimal model of
  gravitino dark matter}},
  \href{https://doi.org/10.1103/PhysRevD.95.095002}{\emph{Phys. Rev. D}
  {\bfseries 95} (2017) 095002}
  [\href{https://arxiv.org/abs/1701.06574}{{\ttfamily 1701.06574}}].

\bibitem{Dudas:2017rpa}
E.~Dudas, Y.~Mambrini and K.~Olive, \emph{{Case for an EeV Gravitino}},
  \href{https://doi.org/10.1103/PhysRevLett.119.051801}{\emph{Phys. Rev. Lett.}
  {\bfseries 119} (2017) 051801}
  [\href{https://arxiv.org/abs/1704.03008}{{\ttfamily 1704.03008}}].

\bibitem{Bhattacharyya:2018evo}
G.~Bhattacharyya, M.~Dutra, Y.~Mambrini and M.~Pierre, \emph{{Freezing-in dark
  matter through a heavy invisible Z'}},
  \href{https://doi.org/10.1103/PhysRevD.98.035038}{\emph{Phys. Rev. D}
  {\bfseries 98} (2018) 035038}
  [\href{https://arxiv.org/abs/1806.00016}{{\ttfamily 1806.00016}}].

\bibitem{Bernal:2018qlk}
N.~Bernal, M.~Dutra, Y.~Mambrini, K.~Olive, M.~Peloso and M.~Pierre,
  \emph{{Spin-2 Portal Dark Matter}},
  \href{https://doi.org/10.1103/PhysRevD.97.115020}{\emph{Phys. Rev. D}
  {\bfseries 97} (2018) 115020}
  [\href{https://arxiv.org/abs/1803.01866}{{\ttfamily 1803.01866}}].

\bibitem{Dudas:2018npp}
E.~Dudas, T.~Gherghetta, K.~Kaneta, Y.~Mambrini and K.A.~Olive,
  \emph{{Gravitino decay in high scale supersymmetry with R -parity
  violation}}, \href{https://doi.org/10.1103/PhysRevD.98.015030}{\emph{Phys.
  Rev. D} {\bfseries 98} (2018) 015030}
  [\href{https://arxiv.org/abs/1805.07342}{{\ttfamily 1805.07342}}].

\bibitem{Barman:2022tzk}
B.~Barman, N.~Bernal, Y.~Xu and O.~Zapata, \emph{{Ultraviolet freeze-in with a
  time-dependent inflaton decay}},
  \href{https://doi.org/10.1088/1475-7516/2022/07/019}{\emph{JCAP} {\bfseries
  07} (2022) 019} [\href{https://arxiv.org/abs/2202.12906}{{\ttfamily
  2202.12906}}].

\bibitem{Turner:1983he}
M.S.~Turner, \emph{{Coherent Scalar Field Oscillations in an Expanding
  Universe}}, \href{https://doi.org/10.1103/PhysRevD.28.1243}{\emph{Phys. Rev.
  D} {\bfseries 28} (1983) 1243}.

\bibitem{Ichikawa:2008ne}
K.~Ichikawa, T.~Suyama, T.~Takahashi and M.~Yamaguchi, \emph{{Primordial
  Curvature Fluctuation and Its Non-Gaussianity in Models with Modulated
  Reheating}}, \href{https://doi.org/10.1103/PhysRevD.78.063545}{\emph{Phys.
  Rev. D} {\bfseries 78} (2008) 063545}
  [\href{https://arxiv.org/abs/0807.3988}{{\ttfamily 0807.3988}}].

\bibitem{Kofman:1997yn}
L.~Kofman, A.D.~Linde and A.A.~Starobinsky, \emph{{Towards the theory of
  reheating after inflation}},
  \href{https://doi.org/10.1103/PhysRevD.56.3258}{\emph{Phys. Rev. D}
  {\bfseries 56} (1997) 3258}
  [\href{https://arxiv.org/abs/hep-ph/9704452}{{\ttfamily hep-ph/9704452}}].

\bibitem{Brandenberger:1997yf}
R.H.~Brandenberger, \emph{{Inflation and the theory of cosmological
  perturbations}},  in \emph{{ICTP Summer School in High-Energy Physics and
  Cosmology}}, pp.~412--445, 6, 1997
  [\href{https://arxiv.org/abs/astro-ph/9711106}{{\ttfamily
  astro-ph/9711106}}].

\bibitem{Bassett:2005xm}
B.A.~Bassett, S.~Tsujikawa and D.~Wands, \emph{{Inflation dynamics and
  reheating}}, \href{https://doi.org/10.1103/RevModPhys.78.537}{\emph{Rev. Mod.
  Phys.} {\bfseries 78} (2006) 537}
  [\href{https://arxiv.org/abs/astro-ph/0507632}{{\ttfamily
  astro-ph/0507632}}].

\bibitem{Allahverdi:2000ss}
R.~Allahverdi, \emph{{Thermalization after inflation and reheating
  temperature}}, \href{https://doi.org/10.1103/PhysRevD.62.063509}{\emph{Phys.
  Rev. D} {\bfseries 62} (2000) 063509}
  [\href{https://arxiv.org/abs/hep-ph/0004035}{{\ttfamily hep-ph/0004035}}].

\bibitem{Kofman:1996mv}
L.A.~Kofman, \emph{{The Origin of matter in the universe: Reheating after
  inflation}},  5, 1996
  [\href{https://arxiv.org/abs/astro-ph/9605155}{{\ttfamily
  astro-ph/9605155}}].

\bibitem{Shtanov:1994ce}
Y.~Shtanov, J.H.~Traschen and R.H.~Brandenberger, \emph{{Universe reheating
  after inflation}},
  \href{https://doi.org/10.1103/PhysRevD.51.5438}{\emph{Phys. Rev. D}
  {\bfseries 51} (1995) 5438}
  [\href{https://arxiv.org/abs/hep-ph/9407247}{{\ttfamily hep-ph/9407247}}].

\bibitem{Ahmed:2021fvt}
A.~Ahmed, B.~Grzadkowski and A.~Socha, \emph{{Implications of time-dependent
  inflaton decay on reheating and dark matter production}},
  \href{https://doi.org/10.1016/j.physletb.2022.137201}{\emph{Phys. Lett. B}
  {\bfseries 831} (2022) 137201}
  [\href{https://arxiv.org/abs/2111.06065}{{\ttfamily 2111.06065}}].

\bibitem{Banerjee:2022fiw}
A.~Banerjee and D.~Chowdhury, \emph{{Fingerprints of freeze-in dark matter in
  an early matter-dominated era}},
  \href{https://doi.org/10.21468/SciPostPhys.13.2.022}{\emph{SciPost Phys.}
  {\bfseries 13} (2022) 022}
  [\href{https://arxiv.org/abs/2204.03670}{{\ttfamily 2204.03670}}].

\bibitem{Affleck:1984fy}
I.~Affleck and M.~Dine, \emph{{A New Mechanism for Baryogenesis}},
  \href{https://doi.org/10.1016/0550-3213(85)90021-5}{\emph{Nucl. Phys. B}
  {\bfseries 249} (1985) 361}.

\bibitem{Dine:1995kz}
M.~Dine, L.~Randall and S.D.~Thomas, \emph{{Baryogenesis from flat directions
  of the supersymmetric standard model}},
  \href{https://doi.org/10.1016/0550-3213(95)00538-2}{\emph{Nucl. Phys. B}
  {\bfseries 458} (1996) 291}
  [\href{https://arxiv.org/abs/hep-ph/9507453}{{\ttfamily hep-ph/9507453}}].

\bibitem{Kofman:2003nx}
L.~Kofman, \emph{{Probing string theory with modulated cosmological
  fluctuations}},  \href{https://arxiv.org/abs/astro-ph/0303614}{{\ttfamily
  astro-ph/0303614}}.

\bibitem{Fujita:2016vfj}
T.~Fujita and K.~Harigaya, \emph{{Hubble induced mass after inflation in
  spectator field models}},
  \href{https://doi.org/10.1088/1475-7516/2016/12/014}{\emph{JCAP} {\bfseries
  12} (2016) 014} [\href{https://arxiv.org/abs/1607.07058}{{\ttfamily
  1607.07058}}].

\bibitem{Karam:2020skk}
A.~Karam, T.~Markkanen, L.~Marzola, S.~Nurmi, M.~Raidal and A.~Rajantie,
  \emph{{Novel mechanism for primordial perturbations in minimal extensions of
  the Standard Model}},
  \href{https://doi.org/10.1007/JHEP11(2020)153}{\emph{JHEP} {\bfseries 11}
  (2020) 153} [\href{https://arxiv.org/abs/2006.14404}{{\ttfamily
  2006.14404}}].

\bibitem{Co:2019jts}
R.T.~Co, L.J.~Hall and K.~Harigaya, \emph{{Axion Kinetic Misalignment
  Mechanism}},
  \href{https://doi.org/10.1103/PhysRevLett.124.251802}{\emph{Phys. Rev. Lett.}
  {\bfseries 124} (2020) 251802}
  [\href{https://arxiv.org/abs/1910.14152}{{\ttfamily 1910.14152}}].

\bibitem{Co:2020dya}
R.T.~Co, L.J.~Hall, K.~Harigaya, K.A.~Olive and S.~Verner, \emph{{Axion Kinetic
  Misalignment and Parametric Resonance from Inflation}},
  \href{https://doi.org/10.1088/1475-7516/2020/08/036}{\emph{JCAP} {\bfseries
  08} (2020) 036} [\href{https://arxiv.org/abs/2004.00629}{{\ttfamily
  2004.00629}}].

\bibitem{Hannestad:2004px}
S.~Hannestad, \emph{{What is the lowest possible reheating temperature?}},
  \href{https://doi.org/10.1103/PhysRevD.70.043506}{\emph{Phys. Rev. D}
  {\bfseries 70} (2004) 043506}
  [\href{https://arxiv.org/abs/astro-ph/0403291}{{\ttfamily
  astro-ph/0403291}}].

\bibitem{Ballesteros:2020adh}
G.~Ballesteros, M.A.G.~Garcia and M.~Pierre, \emph{{How warm are non-thermal
  relics? Lyman-$\alpha$ bounds on out-of-equilibrium dark matter}},
  \href{https://doi.org/10.1088/1475-7516/2021/03/101}{\emph{JCAP} {\bfseries
  03} (2021) 101} [\href{https://arxiv.org/abs/2011.13458}{{\ttfamily
  2011.13458}}].

\bibitem{Landau:1953um}
L.D.~Landau and I.~Pomeranchuk, \emph{{Limits of applicability of the theory of
  bremsstrahlung electrons and pair production at high-energies}}, {\emph{Dokl.
  Akad. Nauk Ser. Fiz.} {\bfseries 92} (1953) 535}.

\bibitem{Landau:1953gr}
L.D.~Landau and I.~Pomeranchuk, \emph{{Electron cascade process at very
  high-energies}}, {\emph{Dokl. Akad. Nauk Ser. Fiz.} {\bfseries 92} (1953)
  735}.

\bibitem{Migdal:1955nv}
A.B.~Migdal, \emph{{[Kvantovoe kineticheskoe uravnenie dlya mnogokratnogo
  rasseyaniya]}}, {\emph{Dokl. Akad. Nauk SSSR} {\bfseries 105} (1955) 77}.

\bibitem{Migdal:1956tc}
A.B.~Migdal, \emph{{Bremsstrahlung and pair production in condensed media at
  high-energies}}, \href{https://doi.org/10.1103/PhysRev.103.1811}{\emph{Phys.
  Rev.} {\bfseries 103} (1956) 1811}.

\bibitem{Soni:2016gzf}
A.~Soni and Y.~Zhang, \emph{{Hidden SU(N) Glueball Dark Matter}},
  \href{https://doi.org/10.1103/PhysRevD.93.115025}{\emph{Phys. Rev. D}
  {\bfseries 93} (2016) 115025}
  [\href{https://arxiv.org/abs/1602.00714}{{\ttfamily 1602.00714}}].

\bibitem{Kolb:1990vq}
E.W.~Kolb and M.S.~Turner, \emph{{The Early Universe}}, vol.~69 (1990),
  \href{https://doi.org/10.1201/9780429492860}{10.1201/9780429492860}.

\bibitem{Planck:2018vyg}
{\scshape Planck} collaboration, \emph{{Planck 2018 results. VI. Cosmological
  parameters}},
  \href{https://doi.org/10.1051/0004-6361/201833910}{\emph{Astron. Astrophys.}
  {\bfseries 641} (2020) A6}
  [\href{https://arxiv.org/abs/1807.06209}{{\ttfamily 1807.06209}}].

\end{thebibliography}\endgroup
\end{document}